%% file: Paper.tex
\documentclass[usenatbib, english]{mn2e}
\usepackage[fleqn]{amsmath}
\usepackage{array,multirow,graphicx}
\usepackage{txfonts}
\usepackage{lmodern}
\bibpunct{(}{)}{;}{a}{}{,}
\usepackage{floatrow}
\usepackage{array}
\usepackage{relsize}
\usepackage[font=small,labelfont=bf]{caption}
\usepackage{bm}

\usepackage{listings}
\newfloatcommand{capbtabbox}{table}[][\FBwidth]

\def\app#1#2{%
\mathrel{%
\setbox0=\hbox{$#1\sim$}%
\setbox2=\hbox{%
\rlap{\hbox{$#1\propto$}}%
\lower1.1\ht0\box0%
}%
\raise0.25\ht2\box2%
}%
}

\title[The \textsc{Dark Sage} semi-analytic model]{Building disc structure and galaxy properties through angular momentum: The {\sc Dark Sage} semi-analytic model}
\author[A.~R.~H.~Stevens et al.]{Adam R.~H.~Stevens,$^1$\thanks{E-mail: astevens@astro.swin.edu.au} Darren J.~Croton$^1$ and Simon J.~Mutch$^2$\\
$^1$Centre for Astrophysics \& Supercomputing, Swinburne University of Technology, Hawthorn, VIC 3122, Australia\\
$^2$School of Physics, The University of Melbourne, Parkville, VIC 3010, Australia}

\begin{document}

\pagerange{\pageref{firstpage}--\pageref{lastpage}} \pubyear{2016}

\maketitle

\label{firstpage}

\begin{abstract}
We present the new semi-analytic model of galaxy evolution, \textsc{Dark Sage}, a heavily modified version of the publicly available \textsc{sage} code. The model is designed for detailed evolution of galactic discs.  We evolve discs in a series of annuli with fixed specific angular momentum, which allows us to make predictions for the radial and angular-momentum structure of galaxies.  Most physical processes, including all channels of star formation and associated feedback, are performed in these annuli.  We present the surface density profiles of our model spiral galaxies, both as a function of radius and specific angular momentum, and find the discs naturally build a pseduobulge-like component. Our main results are focussed on predictions relating to the integrated mass--specific angular momentum relation of stellar discs.  The model produces a distinct sequence between these properties in remarkable agreement with recent observational literature.  We investigate the impact Toomre disc instabilities have on shaping this sequence and find they are crucial for regulating both the mass and spin of discs.  Without instabilities, high-mass discs would be systematically deficient in specific angular momentum by a factor of $\sim$2.5, with increased scatter.  Instabilities also appear to drive the direction in which the mass--spin sequence of spiral galaxy discs evolves.  With them, we find galaxies of fixed mass have higher specific angular momentum at later epochs.
\end{abstract}


\input{Sec1}

\input{Sec2}

\input{Sec3}

\input{Sec5}

\input{Sec6}

\input{Sec8}

\section*{Acknowledgements}
ARHS thanks Chiara Tonini, Karl Glazebrook, Danail Obreschkow, Toby Brown, Gregory Poole, Katherine Mack, Marie Martig, Rebecca Allen, Jennifer Piscionere, Manodeep Sinha, and Gabriele Pezzulli for conversations and advice relating to this work.  We thank the referee for a prompt and considerate report of this paper.  This work made use of the \textsc{sage} codebase, which is publicly available at https://github.com/darrencroton/sage~.

\input{Bibliography}
\appendix
\input{Sec4}
\input{AppA}
\end{document}

%% file: Sec1.tex
\section{Introduction}
\label{sec:intro}

Efforts to develop a comprehensive theory of galaxy formation and evolution have gained significant momentum over the last few decades.  Cosmological $N$-body simulations have not only shown us how the large-scale structure of the Universe forms, but also how overdensities give rise to haloes, the formation sites of galaxies.  Under the generally accepted $\Lambda$CDM paradigm, these haloes merge throughout time, building the Universe hierarchically.  In 30 years of running these simulations, the community has improved the level of detail, i.e.~the number of particles, by over seven orders of magnitude \citep[cf.][]{davis85,skillman14}.  Yet we are still striving for a complete theoretical picture of galaxy evolution in this framework.

One piece to this puzzle lies in our understanding of the angular momentum of galaxies. Specific angular momentum, otherwise referred to as `spin', is arguably one of the most fundamental properties of a galaxy.  Following \citet{fall83}, there has been a recent focus in the literature on how the spin of a galaxy is correlated with its mass and related to its morphology \citep[e.g.][]{rf12,fr13,og14,genel15,teklu15,zavala16}.  These findings suggest one should be able to describe galaxy evolution \emph{explicitly} as a function of spin.  Indeed, several efforts have been made in the past to do this \citep[e.g.][]{dalcanton97,firmani00,bosch01,stringer07}.  To truly test this idea requires numerical, cosmological simulations in which galaxies are evolved with a full consideration of all relevant astrophysics.

Two modes of approach for evolving galaxies in a cosmological framework in supercomputers are prevalent in the literature.  Hydrodynamic cosmological simulations are the most direct approach, where baryonic particles, with all their detailed `small-scale' physics, are co-evolved with the dark matter.  These are now capable of reproducing the array of galaxy types observed in the Universe, with properties not unlike real galaxies \citep[e.g.][]{illustris,eagle}.  The alternative method is the semi-analytic model, which treats the evolution of baryons, and hence galaxies, as a post-processing step of an $N$-body simulation.  While fine structural detail of galaxies is lost, orders of magnitude of computational efficiency is gained through the semi-analytic approach, allowing for a rapid exploration of the vast parameter space governing galaxy evolution.

With this paper, we present \textsc{Dark Sage}, a semi-analytic model with an updated approach to galactic-disc evolution, explicitly based on angular momentum.  This was developed from our publicly available\footnote{https://github.com/darrencroton/sage} code \textsc{sage} \citep[Semi-Analytic Galaxy Evolution,][]{sage}, but has been vastly altered.  Where \textsc{sage} only evolved the integrated properties of galaxies like a classic semi-analytic model, \textsc{Dark Sage} also evolves the one-dimensional structure of galactic discs.  Given semi-analytic models have historically served as a theoretical counterpart to observational surveys, it is prevalent for models to begin including disc structure in the era of surveys which are returning structurally resolved details of galaxies in large numbers \citep[e.g.][]{atlas3d,cgs,sami,califa,sluggs,manga}.  Similar to the models of \citet{stringer07} and \citet{dutton09}, we discretise disc structure into bins of specific angular momentum.  This deviates from semi-analytic models developed by \citet{fu10,fu13}, who instead bin by radius, where now the angular momentum of material within discs is naturally conserved.  

In fact, semi-analytic models have recently begun to make a concerted effort to evolve the angular momentum of galaxies, but these efforts have typically assumed, and not evolved, the angular-momentum \emph{structure} \citep*[e.g.][]{lagos09,guo11,benson12,padilla14,tonini16}.  Other angular-momentum studies have been conducted in post-processing of the semi-analytic model itself, after the historical properties of the galaxies have been determined \citep[e.g.][]{lagos15}.  \textsc{Dark Sage}'s point of difference is that it \emph{self-consistently} evolves disc structure and the integrated properties of galaxies, inclusive of angular momentum as a vector quantity.

This paper is laid out as follows. In Section \ref{sec:sims} we describe the simulation upon which the model is run.  Section \ref{sec:physics} gives a complete description of the model itself, including the physics and prescriptions that have gone into it.  Our main results are presented in the following two sections, where we use the model to make predictions about how galaxies evolve and their properties at $z=0$, comparing these to observations. Section \ref{sec:profiles} focusses on the surface density profiles of galaxies, both as a function of radius and specific angular momentum.  Section \ref{sec:mj} looks at the integrated specific angular momentum of galaxies, how this scales with mass, and how this compares to observations and other predictions.  Finally, we offer conclusions and discuss future prospects for the model in Section \ref{sec:outlook}.

All results from our model (and compared data) assume (or have been altered to assume) a \citet{chabrier03} stellar initial mass function and a Hubble constant with $h=0.73$.  We also term the ``virial radius'' as the radius at which the internal averaged matter density is 200 times the critical density of the Universe ($M_{\rm vir} = M_{200}$).

%% file: Sec2.tex
\section{Simulation details}
\label{sec:sims}

The predecessor to this model, \textsc{sage} \citep{sage}, was designed to be modular and optimised for several cosmological $N$-body simulations, including Millennium \citep{millennium}, Bolshoi \citep*{bolshoi}, and GiggleZ \citep{gigglez}.  Of these three, the greatest focus was given to Millennium, as this simulation has a rich history with the development of semi-analytic models in the literature, which tend to incrementally build from one another \citep[e.g.][]{bower06,croton06,sage,delucia07,font08,fu10,fu13,guo11}.  For the sake of simplicity, we only present \textsc{Dark Sage} as run on one simulation in this paper.  We chose Millennium because the underlying \textsc{sage} codebase is well tested on this simulation, the data are readily available, and this makes the results of this paper more closely comparable to the multitude of semi-analytic publications that have used this simulation.

The Millennium simulation followed the best-fitting cosmological parameters of the first-year \emph{Wilkinson Microwave Anisotropy Probe} data \citep{wmap1}: $\Omega_M = 0.25$, $\Omega_{\Lambda} = 0.75$, $\Omega_b = 0.045$, $\sigma_8 = 0.9$, $h = 0.73$.  The simulation was run with the \textsc{gadget-2} code \citep{gadget}, using a periodic box length of 500$h^{-1}$ comoving Mpc, with $2160^3$ particles of mass $8.6 \times 10^8 h^{-1}\,{\rm M}_{\odot}$.  Haloes and substructure were subsequently identified with the \textsc{subfind} code \citep{subfind}, and merger trees were constructed with \textsc{l-halotree} \citep[][supplementary information]{millennium}.  Subhaloes required 20 gravitationally bound particles to be included in the merger trees.

%% file: Sec3.tex
\section{Physics and design of the model}
\label{sec:physics}

\textsc{Dark Sage} is based on the architecture of \textsc{sage} \citep{sage} and its predecessor, \citet{croton06}, but is vastly different in terms of how properties within galactic discs are evolved.  In this section, we fully describe the new evolutionary prescriptions of the model, which are founded on angular momentum, and only briefly touch on the aspects unchanged from \textsc{sage}.  Each of the processes outlined in Sections \ref{ssec:cooling}--\ref{ssec:agn} affects the evolution of the angular-momentum structure of galactic discs.  We provide a cartoon realisation of some of these processes in Fig.~\ref{fig:cartoon}.

As is standard for semi-analytic models, many of the prescriptions below include free parameters.  The values of these are compiled in Table \ref{tab:params}.  These were obtained by calibrating the model by hand to a set of observational constraints, focussed on the integrated properties of galaxies in the local Universe.  We only allowed 8 parameters to vary in our calibration.  Our constraints include mass functions for each of stars \citep*{baldry08}, H\,\textsc{i} \citep{zwaan05}, and H$_2$ \citep*{keres03};  the H\,\textsc{i}--stellar mass scaling relation \citep{brown15}; the black hole--bulge relation \citep*{scott13}; the Baryonic Tully--Fisher relation \citep*{stark09}; the mass--metallicity relation of galaxies \citep{tremonti04}; and the mean Universal star formation rate density history \citep*{somerville01}. We present the stellar mass function in Fig.~\ref{fig:smf} and provide further details and plots regarding the calibration process in Appendix \ref{sec:constraints}, where we also compare our calibrations against alternative observational datasets [\citet{martin10} for the H\,\textsc{i} mass function, \citet{obreschkow09} for the H$_2$ mass function, and \citet{madau14} for the star formation history].  For all plots in this paper, including the mass functions, we only present results for systems whose masses of the relevant species are above the median for a (sub)halo with 50 particles (e.g.~$\sim$$10^{8.5}\,{\rm M}_{\odot}$ for stellar mass).  We have found both \textsc{sage} and \textsc{Dark Sage} can reliably find acceptable agreement with observational constraints above this limit, and hence we find this as an appropriate lower limit for our results.

\begin{table*}
	\centering
	\begin{tabular}{l l l l c} \hline
		Parameter & Description & Value & Fixed & Section\\ \hline
		$f_b$ & Cosmic baryon fraction & 0.17 & Yes & -- \\
		$\epsilon_{\rm SF}$ & Star formation efficiency from H$_2$ [$10^{-4}$ Myr$^{-1}$] & 3.96 & No & \ref{ssec:sf} \\
		$\theta_{\rm thresh}$ & Threshold angle for stars and gas to be considered coplanar [degrees] & 10.0 & Yes & \ref{ssec:sf} \\
		$Y$ & Yield of metals from new stars & 0.025 & No & \ref{ssec:sf} \\
		$\mathcal{R}$ & Instantaneous recycling fraction & 0.43 & Yes & \ref{ssec:sf} \\
		$\epsilon_{\rm disc}$ & Mass-loading factor due to supernovae at $\Sigma_{0,\rm gas}$ & 6.0 & No & \ref{ssec:sn} \\
		$\Sigma_{0,\rm gas}$ & Surface density scaling for supernova reheating [$\mathrm{M}_{\odot}\, \mathrm{pc}^{-2}$] & 8.0 & No & \ref{ssec:sn} \\ 
		$\epsilon_{\rm halo}$ & Efficiency of supernovae to unbind gas from the hot halo & 0.4 & No & -- \\
		$\vartheta_t$ & Precession angle of gas discs about stars in a dynamical time [degrees] & 5.0 & Yes & \ref{ssec:precession} \\
		$f_{\rm move}$ & Fraction of unstable gas that moves to adjacent annuli & 0.3 & No & \ref{ssec:instab}\\
		$f_{\rm major}$ & Threshold mass ratio for a major merger & 0.3 & Yes & \ref{ssec:mergers} \\ 
		$f_{\rm BH}$ & Rate of black hole growth during quasar mode accretion & 0.03 & No & \ref{ssec:mergers} \\ 
		$\kappa_{\rm R}$ & Radio mode feedback efficiency & 0.035 & No & -- \\
		$\kappa_{\rm Q}$ & Quasar mode feedback efficiency & 0.005 & Yes & -- \\ \hline
	\end{tabular}
	\caption{Fiducial parameter values for \textsc{Dark Sage} used for all the results presented in this paper.  The fourth column indicates whether the parameters were fixed prior to calibration or considered free during calibration.  The fifth column gives the section in this paper where each parameter is first defined or used mathematically.  See \citet{sage} for the parameters not explicitly discussed in this paper.}
	\label{tab:params}
\end{table*}

\begin{figure}
	\centering
	\includegraphics[width=0.95\textwidth]{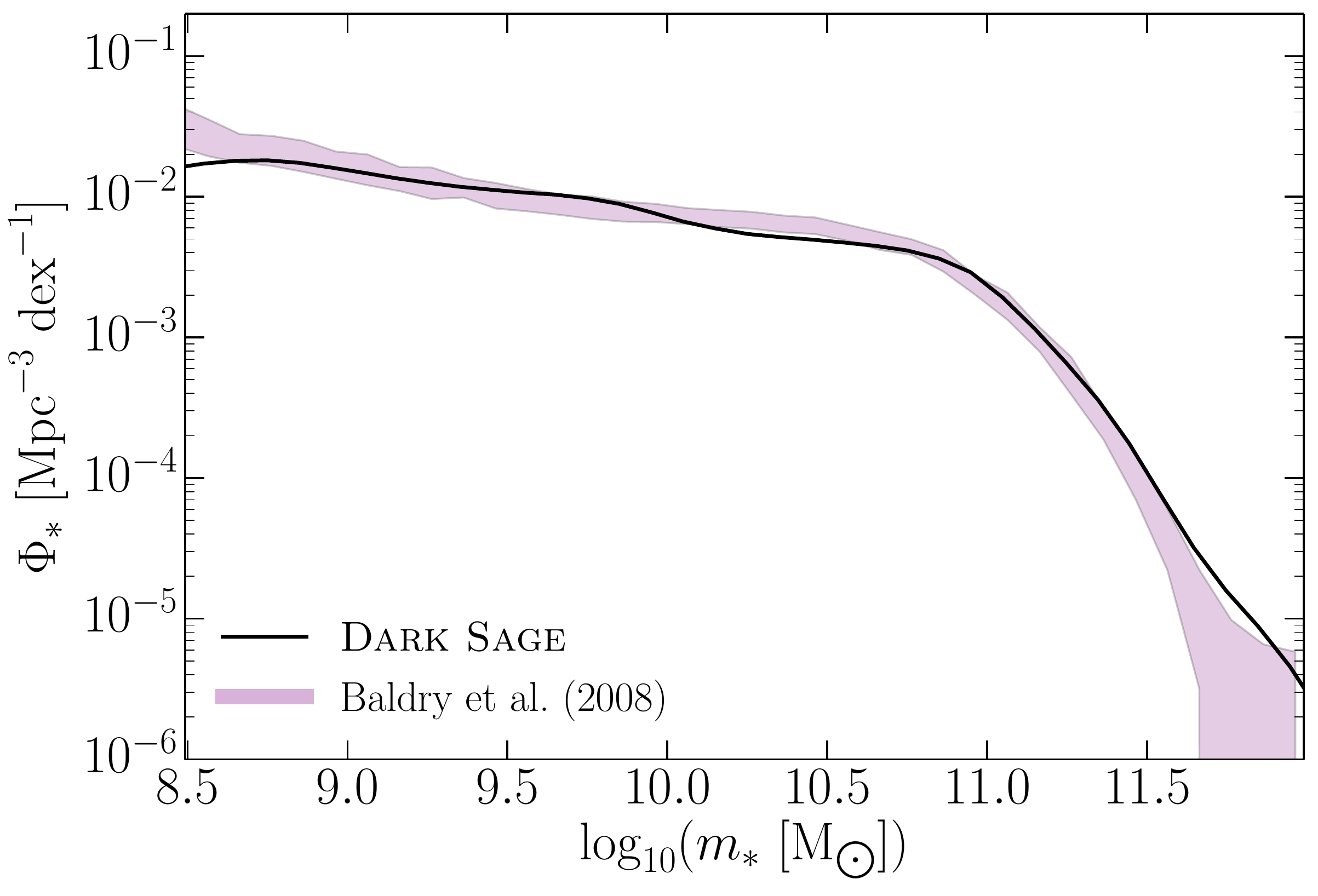}
	\caption{Stellar mass function of \textsc{Dark Sage} galaxies at $z=0$.  The compared observational data of \citet{baldry08} were used to calibrate the model.}
	\label{fig:smf}
\end{figure}

\subsection{Baryonic reservoirs and disc structure}
\label{ssec:reservoirs}

Akin to \textsc{sage} and other popular semi-analytic models, every halo from the simulation \textsc{Dark Sage} is run on is given singular reservoirs for hot gas, ejected gas (hot gas unavailable for cooling), a black hole mass, a merger-driven bulge, an instability-driven bulge, and intracluster stars.  Subhaloes (where satellite galaxies live) include these reservoirs except for the ejected gas and intracluster stars. Discs are instead broken into 30 reservoirs, allowing the one-dimensional structure of discs to be directly evolved within the hierarchical framework.  The processes by which these reservoirs grow and how baryons move between them will be described in the following subsections.

We draw inspiration from the \citet{fu10,fu13} models for breaking galactic discs into annuli, but make several important points of difference.  First, instead of breaking discs into bins of radius, $r$, we break discs into bins of specific angular momentum, 
\begin{equation}
j = r\, v_{\rm circ}(r)~,
\label{eq:j}
\end{equation}
assuming baryons in discs follow Keplerian orbits [see Appendix \ref{app:rotcurves} for how $v_{\rm circ}(r)$ is calculated].  This binning method, also used in the models of \citet{stringer07} and \citet{dutton09}, means \emph{we evolve galactic discs explicitly as a function of specific angular momentum}.  Because $j$ increases monotonically with radius, each bin still represents a disc annulus.  The primary advantage of binning this way is that we naturally conserve angular momentum within discs, something that does not happen for radial bins, as the velocity structure of galaxies is very dynamic.  Thus, we do not need to manually account for radial flows of gas, e.g.~as done by \citet{fu13}.   

\begin{figure*}
	\centering
	\includegraphics[width=0.95\textwidth]{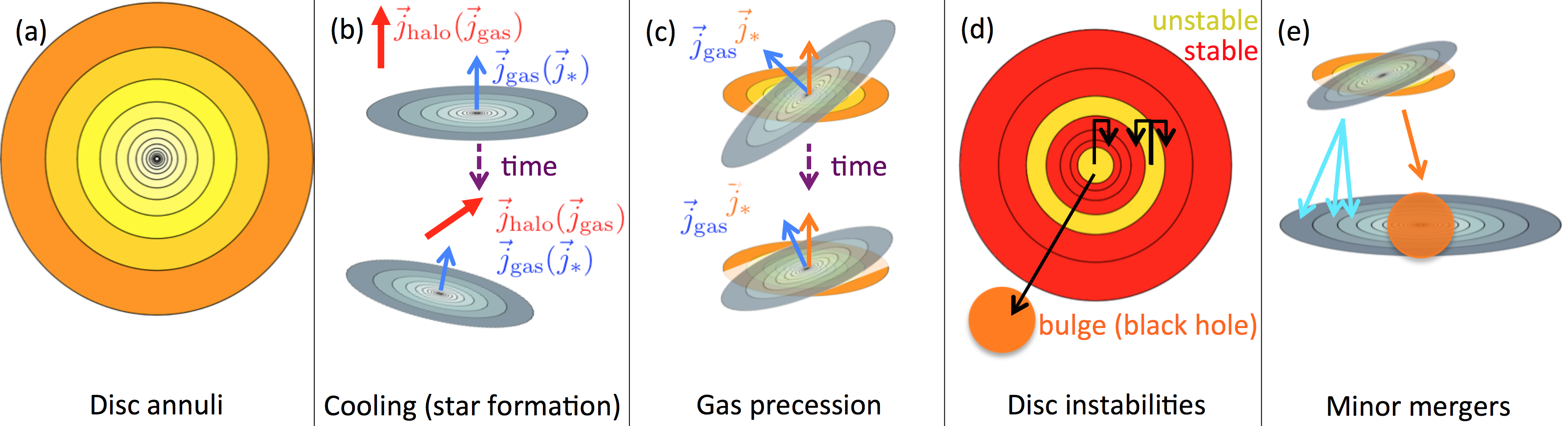}
	\caption{Cartoon of the \textsc{Dark Sage} semi-analytic model.  \emph{Panel (a)}: Galactic discs are broken into annuli, with constant density assumed within a given annulus.  Galaxy evolution processes take place in these annuli (see Section \ref{ssec:reservoirs}).  \emph{Panel (b)}:  Gas discs cools with a spin vector parallel to that of the halo.  Subsequent cooling episodes may occur when the halo and disc spin vectors have an angular offset, leading to a change in the disc's vector's direction (see Section \ref{ssec:cooling}).  Star formation is an analogous process, where the spin of the gas disc is used for new star formation episodes, which can lead to offsets in the gas and stellar discs (see Section \ref{ssec:sf}).  \emph{Panel (c)}: The gas disc precesses about the stellar spin axis, maintaining coplanarity (see Section \ref{ssec:precession}).  \emph{Panel (d)}: Unstable disc annuli transfer some of their mass to adjacent annuli, conserving angular momentum in the process.  For the innermost annulus, unstable stars are partially transferred to the bulge, while unstable gas feeds the black hole (see Section \ref{ssec:instab}).  \emph{Panel (e)}: Merging satellite galaxies have their stellar content shifted to the bulge of the central galaxy, while their gas goes to disc annuli of the appropriate specific angular momentum (see Section \ref{ssec:minor}).  Further processes are described throughout Section \ref{sec:physics}.}
	\label{fig:cartoon}
\end{figure*}

To first order, one would expect the surface density of discs to fall exponentially with radius and for $j$ to increase linearly with radius (but see Section \ref{ssec:discprofs} and Appendix \ref{app:rotcurves}).  As such, we use fixed specific-angular-momentum bins for all galaxies, spaced equally in logarithmic space.  More specifically, the outer edge of each bin is given by
\begin{equation}
j_{i} = j_1 \times f_{\rm step}^{i-1}~,~~~i = 1,2,...,30~
\end{equation}
(with the inner edge of the first annulus at $j=0$).  Our choice of using 30 annuli per galaxy allows for sufficient resolution to evolve disc structure accurately and follows \citet{fu10}, but truthfully is arbitrary (\citealt{fu13} suggest as few as 12 annuli might be sufficient for producing galaxies with realistic properties).  We find our results are converged with $j_1 = 1.0h^{-1}\,\mathrm{kpc}\,\mathrm{km}\,\mathrm{s}^{-1}$, so we adopt this value.  $f_{\rm step}$ is then set so the outermost annulus roughly reaches the virial radius of a Milky Way-like system.  This gives us $f_{\rm step} = 1.4$.  For an image of an example disc, see Fig.~\ref{fig:cartoon}a.

The second difference from the \citeauthor{fu10} models is that we do not force disc stars and gas in the galaxy to be coplanar.  Instead we track separate spin vectors for each disc component \citep[similar to][]{lagos15}, but ensure they do not persist with non-physical, large angular separations through a gas precession module (Section \ref{ssec:precession}).  Hereafter, we shall refer to these components as the gas disc and stellar disc.

\subsection{Gas infall and dissipation}
\label{ssec:infall}

As haloes accumulate mass from their surroundings, so too do they accumulate gas.  Equally, if haloes lose mass, some hot gas should also be lost.  We follow the same prescription as \textsc{sage} for determining the rate at which hot gas falls in or dissipates from haloes, which is based on the idea of maintaining the cosmic baryon fraction within haloes.  This is modulated by a reionization module, which accounts for heating from ionizing photons leading to a decrease in baryon densities at high redshift \citep{efstathiou92,gnedin00}.  We refer the reader to sections 4 and 5 of \citet{sage} for full details.


\subsection{Cooling of hot gas and formation of the interstellar medium}
\label{ssec:cooling}

Galaxies initially form and grow from radiative cooling and condensation of hot gas in haloes \citep{white78}.  We follow the prescription of \textsc{sage} for determining how much gas cools at each time interval in the model.  Based on \citet{white91}, the hot gas is assumed to be a singular isothermal sphere at the halo's virial temperature.  The gas cooling rate then follows the similarity solutions of \citet{bertschinger89}, which is limited by the free-fall time-scale of the halo, and uses the cooling function tables from \citet{sutherland93}, which depend on temperature and metallicity.  For further details, see section 3.2.1 of \citet{croton06} and section 6 of \citet{sage}.

Over the time step, $\Delta t$, the amount of cooling gas, $\Delta m_{\rm cold}$, must be spread out to form or add to a galactic disc.  Following the disc formation scenario proposed by \citet{fall80}, it was shown explicitly by \citet{fall83} and \citet*{mo98} that if the gas from an isothermal sphere cools into a disc with an exponential surface density profile, 
\begin{equation}
\Sigma_{\rm cool}(r) = \frac{\Delta m_{\rm hot \rightarrow cold}}{2 \pi r_d^2} e^{-r/r_d}~,
\label{eq:Sigma_cool}
\end{equation}
the specific angular momentum of the gas will be conserved if the newly formed, centrifugally supported disc rotates with a constant circular velocity, $v_{\rm circ} = V_{\rm vir}$, and has a scale radius of
\begin{equation}
r_d = \frac{\lambda}{\sqrt{2}} R_{\rm vir}~.
\label{eq:r_d}
\end{equation}
Here, $\lambda$ is the spin parameter of the halo:
\begin{equation}
\lambda \equiv \frac{J |E|^{1/2}}{G M_{\rm vir}^{5/2}} = \frac{j_{\rm halo}}{\sqrt{2} V_{\rm vir} R_{\rm vir}}~,
\label{eq:lambda}
\end{equation}
where $J$ and $E$ are the total angular momentum and energy of the halo, respectively \citep{peebles69,bullock01}.  Here, the net specific angular momentum of the cooling gas matches that of the halo, $j_{\rm halo}$.

The above cooling picture can be generalised by considering the distribution of cooling gas as a function of specific angular momentum:
\begin{equation}
\Sigma_{\rm cool}(j) = \frac{\Delta m_{\rm cold}}{2 \pi r_d^2} e^{-j/r_d V_{\rm vir}}~.
\label{eq:cooling}
\end{equation}
We apply equation (\ref{eq:cooling}) when determining how the cooling gas is distributed among the $j$ bins in \textsc{Dark Sage}, thereby ensuring angular momentum is conserved as the gas cools.  By using this instead of equation (\ref{eq:Sigma_cool}), we no longer assume cooling gas to have a singular rotational velocity.  Cooling gas of a given $j$ is then assumed to localise itself with gas in the disc of the same $j$.

The orientation of the cooling gas's angular momentum is always assumed to be parallel to that of the halo at the time of cooling.  As both the magnitude and direction of a halo's spin is very dynamic over a Hubble time, this means subsequent cooling episodes likely occur at angles to one another.  If a gas disc does not exist prior to a cooling episode, then a gas disc is initialised with a spin vector parallel to the halo.  If instead a gas disc already exists, the angular-momentum vectors of the pre-existing and cooling gas discs are summed together to define the new orientation of the gas disc.  Both the pre-existing and cooling gas discs are then projected onto this new plane, where any orthogonal component of angular momentum is assumed to dissipate, and the discs are summed together.  Discs are therefore more compact than in \textsc{sage}. See Fig.~\ref{fig:cartoon}b for a depiction of this process.

It is worth noting that neither the direction nor magnitude of a halo's spin from a cosmological simulation is totally precise \citep[see, e.g.,][]{bullock01,bett10}.  The relative uncertainty of $j_{\rm halo}$ is correlated with both the number of particles in the halo and the magnitude of $j_{\rm halo}$ itself (Contreras et al. in preparation).  A careful qualitative analysis of these uncertainties are left to be presented by Contreras et al., but we do not expect this to significantly impact the results of semi-analytic models that are primarily concerned with galaxies in well-resolved haloes, where uncertainties in angular momentum are minimised.


\subsection{Passive star formation from molecular gas}
\label{ssec:sf}

In a gas disc, fragmentation leads to the formation of giant molecular clouds, which in turn collapse to form stars.  At each time-step, we allow stars to form in each annulus based on the local H$_2$ content.  Specifically, the star formation rate surface density is
\begin{equation}
\Sigma_{\rm SFR}(r) = \epsilon_{\rm SF}\, \Sigma_{\rm H_2}(r)~,
\end{equation}
where $\epsilon_{\rm SF}$ is the star formation efficiency.  This relation is supported observationally \citep[e.g.~from the inner parts of spiral galaxies from][]{leroy08} and has proven successful in semi-analytic models previously \citep[e.g.][]{fu13}.

To form stars with this model, the H$_2$ content in each bin must first be calculated.  We follow \citet{blitz04} and assume that the mid-plane pressure of the gas determines the ratio of molecular to atomic hydrogen:
\begin{equation}
R_{\rm H_2}(r) \equiv \frac{\Sigma_{\rm H_2}(r)}{\Sigma_{\rm H\,\textsc{i}}(r)} = \left[\frac{P(r)}{P_0}\right]^{\mathcal{X}}~,
\end{equation}
with $P_0 = 5.93 \times 10^{-13}\, h^2\, {\rm Pa}$ and $\mathcal{X} = 0.92$ \citep[cf.][]{blitz06}.  Following \citet{elmegreen89}, the mid-plane pressure is calculated as
\begin{equation}
P(r) = \frac{\pi}{2}\, G\, \Sigma_{\rm gas}(r) \left[\Sigma_{\rm gas}(r) + \frac{\sigma_{\rm gas}}{\sigma_*(r)} \Sigma_*(r) \right]~,
\label{eq:pressure}
\end{equation}
where $\sigma$ is the velocity dispersion for the subscripted component.  Equation (\ref{eq:pressure}) is only directly applied if the gas and stellar discs are sufficiently aligned.  If their angular separation exceeds some threshold, $\theta_{\rm thresh}$ (nominally $10^{\circ}$), then the expression simplifies to
\begin{equation}
P(r) = \frac{\pi}{2}\, G\, \Sigma^2_{\rm gas}(r)~.
\end{equation}

We assume $\sigma_{\rm gas} = 11$ km\,s$^{-1}$ for all galaxies at all radii \citep{leroy08}.  In truth, the H\,\textsc{i} velocity dispersions of observed spiral galaxies are anti-correlated with radius, but a value of 11 km\,s$^{-1}$ is typical at an optical radius and is generally within a factor of 2 across the entire disc (\citealt{tamburro09}; also see \citealt{zheng13}).  

For the stellar velocity dispersion profile, we follow the relation identified by \citet{bottema93},
\begin{equation}
\sigma_*(r) = \frac{1}{2}\, V_{\rm vir}\, e^{-r / 2r_d}~,
\label{eq:sigma_*}
\end{equation}
where we use $r_d$ from equation (\ref{eq:r_d}).  An exponentially decreasing $\sigma_*(r)$ profile is supported by $N$-body simulations of stellar discs \citep*{khoperskov03} and more recent observations of spiral galaxies \citep{martinsson13}.

Disc gas is composed of hydrogen, helium, and metals.  We also assume some of the gas is in a warm, ionized phase, unavailable for star formation (following \citealt{ob09}, based on Milky Way observations from \citealt{reynolds04}).  As such, we calculate the true fraction of H$_2$ as
\begin{equation}
f_{\rm H_2}(r) \equiv \frac{\Sigma_{\rm H_2}(r)}{\Sigma_{\rm gas}(r)} = \frac{f_{\rm He}\, f_{\rm warm}}{R_{\rm H_2}^{-1}(r) + 1}\, \left[1 -  Z_{\rm gas}(r)\right]~,
\end{equation}
with $f_{\rm He} = 0.75$ and $f_{\rm warm} = 1.3^{-1}$ \citep[as used by][]{fu10},\footnote{While $\Sigma_{\rm gas}$ represents `cold' gas, this still includes the warm, ionized gas in the disc.} where 
\begin{equation}
Z_{\rm gas}(r) \equiv \frac{\Sigma_{Z,\mathrm{gas}}(r)}{\Sigma_{\rm gas}(r)}
\end{equation}
is the metallicity of the gas in the disc.

Every episode of star formation is assumed to form a simple stellar population.  High-mass stars burn bright and die fast, returning gas and newly formed metals to the interstellar medium through stellar winds and supernovae (feedback is described in Section \ref{ssec:sn}).  We use the instantaneous recycling approximation \citep[following][]{cole00} and immediately return a fraction $\mathcal{R}$ of the gas used to form stars in a given annulus back to the gas disc.  The net production of newborn stars is hence described by
\begin{equation}
\Delta \Sigma_{\rm gas \rightarrow *}(r) = (1 - \mathcal{R})\, \Sigma_{\rm SFR}(r)\, \Delta t~.
\end{equation}
The returned gas is also assumed to be enriched with new metals, given by a fixed yield, $Y$ \citep*[as in][]{delucia04}.  The net change in gaseous metals for a given annulus after an episode of star formation is thus
\begin{equation}
\Delta \Sigma_{Z,\mathrm{gas}}(r) = \left\{Y[1-Z_{\rm gas}(r)] - [1-\mathcal{R}] Z_{\rm gas}(r) \right\}  \Sigma_{\rm SFR}(r)\, \Delta t~,
\end{equation}
where the factor of $[1-Z(r)]$ next to $Y$ accounts for the fact that metals cannot produce more metals.

Beyond passive star formation from H$_2$, \textsc{Dark Sage} includes two channels for starbursts (see Sections \ref{ssec:instab} and \ref{ssec:starburst}).  These events follow the same recycling, enrichment, and feedback regimes.

Stars are always born in the plane of the gas disc.  If no stellar disc exists prior to a star formation episode, the stellar disc orientation is initialised as equal to the gas disc's.  If a stellar disc already exists, then the pre-existing and newborn stellar discs are combined in the same manner as gas discs during cooling; the angular-momentum vectors of each disc are summed to define the new stellar disc plane, then both the pre-existing and newborn stellar discs are projected onto this plane, where any orthogonal angular momentum is dissipated (refer again to Fig.~\ref{fig:cartoon}b).

We have tested calculating H$_2$ content with a metallicity-dependent prescription based on \citet{mckee10}.  Using this requires recalibration of the model,  after which our results do not qualitatively change.  We have opted to exclude results from this prescription for the sake of simplicity.


\subsection{Supernova feedback}
\label{ssec:sn}

As the deaths of high-mass stars in the form of Type-II supernovae return mass and metals to the intergalactic medium, so too do they release energy capable of reheating the gas and removing it from the galaxy.  In regions of high density, the energy produced by supernovae is dispersed over a higher mass of gas, thereby heating that gas less and allowing it to cool back faster (while maintaining its angular momentum).  If the time-scales for this are short, then the net amount of heated gas from supernovae should depend on local gas density.  To account for this, we follow the supernova feedback model of \citet{fu10} where the local heating rate surface density of gas is
\begin{equation}
\Sigma_{\rm reheated}(r) \equiv \frac{\Delta \Sigma_{\rm gas \rightarrow hot}(r)}{\Delta t} = \epsilon_{\rm disc} \frac{\Sigma_{0,\mathrm{gas}}}{\Sigma_{\rm gas}(r)} \Sigma_{\rm SFR}(r)~,
\label{eq:sn}
\end{equation}
where $\Sigma_{0,\mathrm{gas}}$ is a reference surface density and $\epsilon_{\rm disc}$ is the mass-loading factor (see Table \ref{tab:params}).  Despite $\Sigma_{\rm gas}(r)$ appearing in the denominator of equation (\ref{eq:sn}), our prescription still preferentially removes low-angular-momentum gas from the disc, as there is simply more of it.  This is consistent with results of hydrodynamic simulations \citep{brook11,brook12}.  

In annuli where $\Sigma_{\rm gas} \ll \Sigma_{\rm 0,gas}$, equation (\ref{eq:sn}) can shut off star formation entirely.  To prevent this, we only use the instantaneous recycling approximation (Section \ref{ssec:sf}) and apply supernova feedback if a star formation event in an annulus would produce $>100 h^{-1}\,\mathrm{M}_{\odot}$ of new stars.  Provided this limit is well below the simulation mass resolution, our results are impervious to its precise value, as stars that form in these annuli contribute little to a galaxy's stellar mass overall.

The energy released by supernovae may be sufficient to not only heat gas out of the disc, but to eject gas from the halo entirely.  To determine both the ejection and reincorporation rate of gas, we follow the standard \textsc{sage} prescription, which is described in section 8.1 of \citet{sage}.


\subsection{Gas precession}
\label{ssec:precession}

While it is nice to not force the gas and stellar discs to be coplanar, it is not physically reasonable (i.e.~not observed) to have a significant number of galaxies where these discs have a large angular offset.\footnote{Offsets between axes of rotation for gas and stars are observed in elliptical galaxies though \cite[e.g.][]{davis11}.}  \citet*{tohline82} discuss that a gas disc embedded in an axisymmetric potential with an initial angular offset will precess to become coaxial or counter-axial with the potential. Hydrodynamic simulations have also shown that a gas disc rotating at an angle to stars should settle to be coaxial or counter-axial within a few galaxy dynamical times if left undisturbed, where the angular-momentum component orthogonal to this axis is lost to other parts of the halo \citep{voort15}.  

Following this narrative, \textsc{Dark Sage} includes a precession prescription for the gas disc.  To ensure a reasonable level of alignment between gas and stars, we assume the potential axis of symmetry to be coaxial with the stars' rotational axis.  Where stellar mass is primarily in a disc or instability-driven bulge, this is coaxial with the stellar disc (see Fig.~\ref{fig:cartoon}c).  Where most stellar mass lies in a merger-driven bulge (i.e.~early-type galaxies), this is coaxial with the bulge's spin (see Section \ref{ssec:mergers}).

We formulate a simple precession rate prescription, which is inversely proportional to a galaxy's dynamical time \citep[inspired by][]{voort15}, determined as a mass-weighted average of the annuli's dynamical times,
\begin{equation}
\frac{\Delta \theta_{\rm gas\,disc}}{\Delta t} = \vartheta_t \left[ \frac{1}{m_{\rm cold}} \sum_{i=1}^{30} m_i \frac{\bar{r}_i}{v_{\rm circ}(\bar{r}_i)} \right]^{-1}~,
\end{equation}
where $\bar{r}_i$ is the average radius of the mass in the $i^{\rm th}$ annulus and $m_i = \Sigma_{\rm gas}(r_i)\, \pi (r_i^2 - r_{i-1}^2)$.  Our precession parameter is fiducially set at $\vartheta_t = 5^{\circ}$ per dynamical time.  The precession angle, $\Delta \theta_{\rm gas\,disc}$, is limited to the angle necessary for gas and stars to become co- or counter-aligned (and hence is always $<90^{\circ}$).

\begin{figure}
	\centering
	\includegraphics[width=0.95\textwidth]{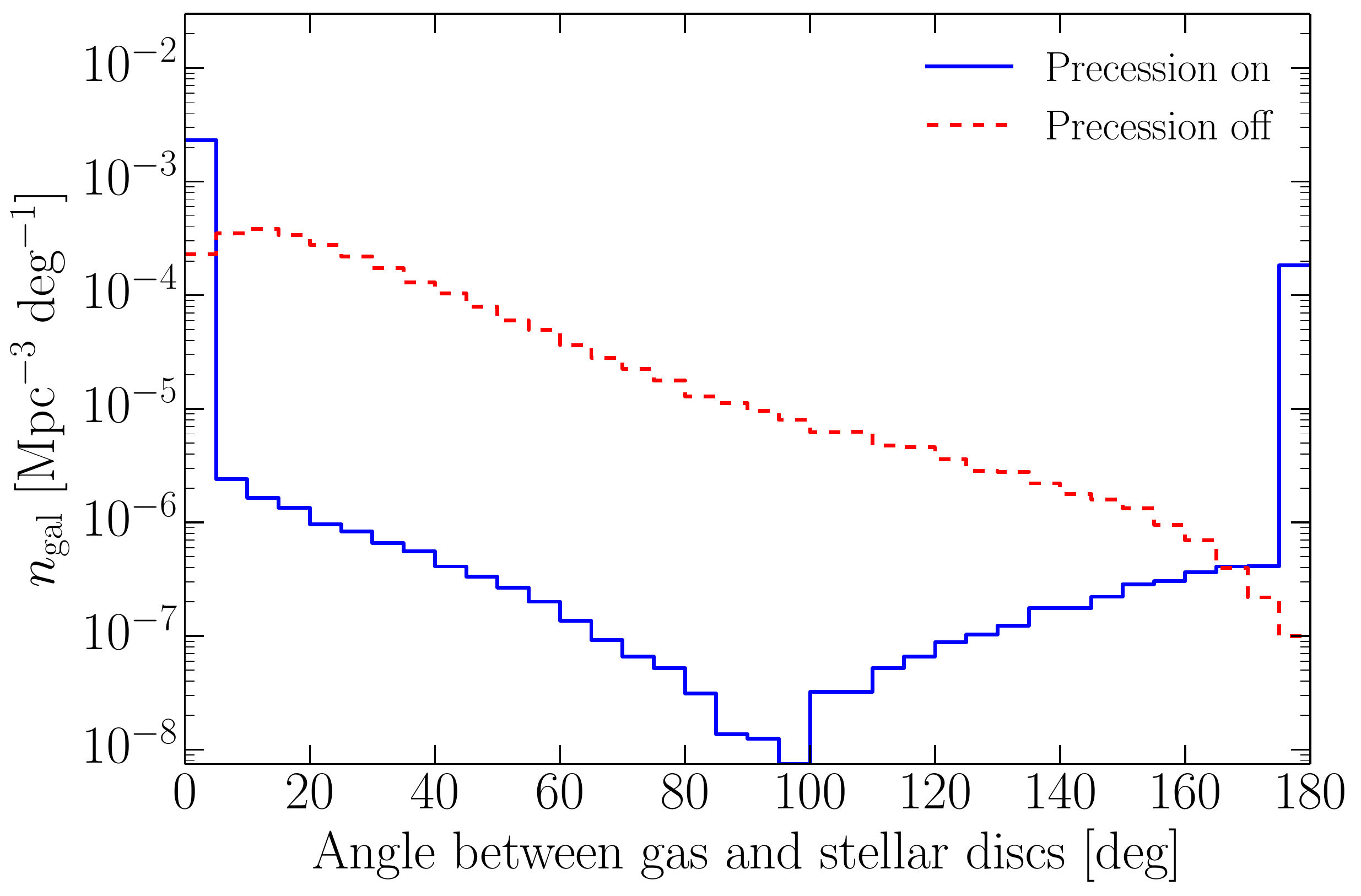}
	\caption{Number density of disc-dominated \textsc{Dark Sage} galaxies (bulge-to-total ratio $< 0.5$) as a function of the angular offset between their gas and stellar discs at $z=0$.  The gas precession module (Section \ref{ssec:precession}) is necessary to obtain galaxies with realistic offsets, where gas and stars are almost always coplanar.}
	\label{fig:offsets}
\end{figure}
	
Fig.~\ref{fig:offsets} shows the importance this module has in the angular offset between stellar and gas discs at redshift zero.  Without it, the number density of disc-dominated galaxies with a particular angular offset declines as a shallow exponential with the offset angle.  With the module, most gas and stellar discs are coplanar by $z=0$.  As expected from the model's design, a small portion (7.25$\%$) of these coplanar systems are counter-aligned.  We do not currently know of any published observational data to compare this against.

The spin direction of a halo changes frequently.  This means the spin direction of the gas disc keeps changing as more cooling happens (Section \ref{ssec:cooling}).  Without precession, the stellar disc will always lag the gas disc when it comes to updating its orientation (new star formation episodes bring the discs more in line -- Section \ref{ssec:sf}).  That lag means it is more likely to have a non-zero (but small) angular offset between the gas and stellar discs, without precession, which is seen by the peak in the dashed distribution in Fig.~\ref{fig:offsets}.  Ultimately, the location of this peak is controlled by the mean rate of change of the haloes' spin directions, which, in this case, comes from Millennium.


\subsection{Disc instabilities}
\label{ssec:instab}

Following cooling episodes (Section \ref{ssec:cooling}) and mergers (Section \ref{ssec:mergers}), we check the level of stability of each annulus.  We calculate a Toomre $Q$ parameter for both the gas and stars for each annulus, where
\begin{equation}
Q_*(r) = \kappa(r)\, \sigma_*(r)\, /\, 3.36\, G\, \Sigma_*(r)
\end{equation}
\citep{toomre64} and
\begin{equation}
Q_{\rm gas}(r) = \kappa(r)\, c_s\, /\, \pi\, G\, \Sigma_{\rm gas}(r)~
\end{equation}
\citep{binney87}, with epicyclic frequency
\begin{equation}
\kappa(r) \equiv \sqrt{\frac{2 v_{\rm circ}(r)}{r^2} \frac{\mathrm{d} j}{\mathrm{d}r} }
\end{equation}
\citep{pringle07}.  For the speed of sound, we approximate $c_s \simeq \sigma_{\rm gas} = 11$ km\,s$^{-1}$. $\sigma_*(r)$ is given by equation (\ref{eq:sigma_*}).  We start by assessing the stability of the outermost annulus of the gas disc and incrementally shift inward.  We then do the same for the stellar disc.  

If the misalignment between the gas and stellar discs exceeds $\theta_{\rm thresh}$, we consider gas and stellar instabilities independently. In that case, should either $Q_*$ or $Q_{\rm gas}$ be $<1$ for an annulus, the unstable mass is calculated as that required to be removed to bring the relevant $Q$ up to $1$.

If the gas and stellar discs are aligned, then stars will affect the stability of gas and vice versa.  We hence calculate a combined $Q_{\rm tot}$ for each annulus, following \citet{romeo11}, where
\begin{subequations}
\begin{equation}
Q_{\rm tot}^{-1}(r) = 
\left\{
\begin{array}{lr}
Q_{\rm gas}^{-1}(r) + W(r)\, Q_*^{-1}(r)\, , & Q_{\rm gas}(r) < Q_*(r) \\
W(r)\, Q_{\rm gas}^{-1}(r) + Q_*^{-1}(r)\, , & Q_{\rm gas}(r) \geq Q_*(r)
\end{array}
\right.
~,
\end{equation}
\begin{equation}
W(r) \equiv \frac{2\, \sigma_{\rm gas}\, \sigma_*(r)}{\sigma_{\rm gas}^2 + \sigma_*^2(r)}~.
\end{equation}
\end{subequations}
If $Q_{\rm tot}<1$ for an annulus, an instability occurs.  In that case, stars and/or gas must be shifted out of that annulus such that $Q_{\rm tot}$ is raised to $1$.  To determine how much of the unstable mass is in the form of stars and gas, we first calculate a value of $Q_{\rm stable}$, where if each of $Q_{\rm gas}$ and $Q_*$ were equal to this value, $Q_{\rm tot}$ would equal $1$:
\begin{equation}
Q_{\rm stable}(r) = 1 + W(r)~.
\end{equation}
If $Q_{\rm gas}>Q_{\rm stable}$, all the unstable mass is in the form of stars.  If $Q_*>Q_{\rm stable}$, all the unstable mass is in the form of gas.  If both are $<Q_{\rm stable}$, first $Q_{\rm gas}$ is raised to $Q_{\rm stable}$, then $Q_*$ is raised as necessary.

Gaseous instabilities can either be resolved by internal motion of gas or by rapid gravitational collapse of the gas to form stars. Once an unstable gas mass has been calculated, we transfer a parametrized fraction, $f_{\rm move}$, of the unstable gas to the adjacent annuli. The proportion of mass moved outward and inward ensures angular momentum is conserved.  For the innermost annulus, the gas that would have moved to a lower-$j$ bin instead feeds the black hole and results in quasar mode feedback (see Section \ref{ssec:agn}).  The remaining gas is consumed in a starburst and associated supernova feedback.  These stars are added to the disc in the same way as an ordinary star formation episode.  After the entire gas disc has been dealt with, we recalculate $Q_*$, then deal with the stellar disc.  This method ensures the discs find stability, where it is not necessary to directly transfer the instability-burst stars to the bulge, as done in \textsc{sage} for example (although, those stars might immediately be seen as unstable and migrate inward anyway).

Unstable stars are simply transferred to the adjacent annuli, again conserving angular momentum.  Should unstable stars exist in the innermost annulus, the portion of unstable stars that lose angular momentum is assumed to lose all of it, and those stars are transferred to the instability-driven bulge.  By construction, the instability-driven bulge has no angular momentum \citep[note, this definition differs to the model of][]{tonini16}.  Fig.~\ref{fig:cartoon}d displays an example case of this process.

Our method of dispersing mass between adjacent annuli of fixed $j$ to resolve gravitational instabilities is complementary to disc models with radial annuli that explicitly impose that annuli torque their neighbours, with the value of that torque dependent on local $Q$ \citep*[e.g.][]{forbes12,forbes14}.  The net intended effect is the same.

Whenever stars or gas shift to an adjacent annulus, they take with them metals in proportion to the metallicity of their original annulus.  This has consequences for the metallicity gradients of galaxies.  This is a subject we intend to look into in a future paper.

An instability is typically regarded as either a global (averaged over the entire disc) or very local process.  Semi-analytic models, by their design, cannot treat truly local instabilities.  While instabilities do not physically happen in annuli, as calculated in \textsc{Dark Sage}, our model does act like a global prescription would.  Because most unstable mass is transferred inward, if annulus $i$ is unstable, the likelihood that annulus $i-1$ will be unstable is raised.  Because we check annulus $i-1$ next, this then cascades all the way to the centre of the galaxy, which has the same external appearance as a global instability.


\subsection{Stripping and accretion of subhaloes}
\label{ssec:subhaloes}

In addition to stripping of hot gas \citep[implemented the same as \textsc{sage} -- see section 10 of][]{sage}, satellite galaxies are subject to ram-pressure stripping of their cold gas, caused by their relative motion to the intergalactic or intracluster medium.  \citet{gunn72} proposed that if ram pressure exceeds the gravitational restoring force per unit area of the galaxy then it will successfully strip the gas.  At each time-step, for each annulus of each satellite, we check if
\begin{equation}
\rho_{\rm hot, cen}(R_{\rm sat})\, v_{\rm sat}^2 \geq 2 \pi G\, \Sigma_{\rm gas}(r) \left[\Sigma_{\rm gas}(r) + \Sigma_*(r) \right] ~,
\label{eq:RPS}
\end{equation}
where $\rho_{\rm hot, cen}$ refers to that of the central galaxy (see equation 3 of \citealt{sage}), $R_{\rm sat}$ is the distance between the central galaxy and satellite, and $v_{\rm sat}$ is the velocity of the satellite relative to the central.  If this criterion is fulfilled, the gas in that annulus is transferred to the hot reservoir of the main halo (associated with the central galaxy).  $\Sigma_*(r)$ is removed from equation (\ref{eq:RPS}) if the gas and stellar discs are not aligned within $\theta_{\rm thresh}$.  Using a combined hydrodynamics and semi-analytic approach, \citet{teece10} show that approximating an isothermal distribution for hot gas in an NFW halo tends to over-predict the strength of ram pressure, especially at higher redshift, but for $z \lesssim 0.5$, this should be accurate within a factor of 2 (see their fig.~3).  We also note that equation (\ref{eq:RPS}) does not consider the orientation of the satellite galaxy.

We do not include orphan galaxies in our model.  Once a subhalo associated with a satellite galaxy is lost, we either merge the galaxy with the central or disrupt it.  Details can be found in section 10 of \citet{sage}.


\subsection{Galaxy mergers}
\label{ssec:mergers}

In line with the standard semi-analytic model narrative, when a galaxy--galaxy merger occurs in the model, we first check the baryonic-mass ratio of the merging galaxies (hot gas and intracluster stars are excluded).  If this is above $f_{\rm major}$ (nominally 0.3), then we class this as a major merger.  Otherwise, fittingly, the merger is classed as minor.  We describe how each type of merger is dealt with below.  Both mergers are dealt with in a similar way to \textsc{sage}, but with extra details regarding discs.

\subsubsection{Major mergers}
\label{ssec:major}
We opt for a simple model of combining gas discs in major mergers.  We sum the angular-momentum vectors of the gas discs of the two systems (as measured relative to their own respective centres of mass), then project both discs onto this axis and sum them together.  All annuli are then subject to a merger starburst (Section \ref{ssec:starburst}).  Some of this gas also directly feeds the black hole and leads to quasar mode feedback.  We follow the phenomenological relation used in \textsc{sage} \citep[from][]{kauffmann00}, but modify it so it applies to each annulus individually.  The growth of the black hole from this channel is hence
\begin{multline}
\Delta m_{\rm cold \rightarrow BH} = f_{\rm BH} \left[1 + \left( \frac{280\,\mathrm{km\,s}^{-1}}{V_{\rm vir}}\right)^2 \right]^{-1} \\ \sum_{i=1}^{30} (m_{i, \rm cen} + m_{i, \rm sat})\, \mathrm{min} \left( \frac{m_{i, \rm sat}}{m_{i, \rm cen}}, \frac{m_{i, \rm cen}}{m_{i, \rm sat}} \right)~,
\label{eq:BHmajor}
\end{multline}
where the $m_i$ values are the contributed gas masses to each annulus from each of the satellite and central.  Because the $j$ distribution of discs, and hence collisional gas in mergers, is weighted toward the low end, the black hole preferentially accretes low-$j$ gas.

The stellar discs and instability-driven bulges of both galaxies are destroyed during a major merger, where all stellar mass forms a merger-driven bulge.  The bulge is assigned a spin direction whose axis is parallel to the orbital axis of the merging galaxies at the last resolved moment before merging.  This axis is only used for gas precession (Section \ref{ssec:precession}) in elliptical galaxies in the model, and thus is not important for our results, which focus on spiral galaxies.

\subsubsection{Minor mergers}
\label{ssec:minor}
For minor mergers, we assume the larger (central) galaxy maintains the integrity of its structure.  If the baryons of the merging satellite are to be added to the disc of the central galaxy, they must be assigned some specific angular momentum relative to the central's centre of mass.  First, we measure the specific angular momentum of the satellite itself, relative to the central, $\vec{j}_{\rm sat} = \vec{R}_{\rm sat}\times \vec{v}_{\rm sat}$, at the last snapshot it is resolved before merging.\footnote{In reality, dynamical friction could reduce the specific angular momentum of a merging satellite between the snapshot of last resolution and the actual time of merging.  In this sense, our calculation of $j_{\rm sat}$ is more of an upper limit.}  We then project this vector onto the central's gas rotation axis.  We then convolve this $j$ with a top-hat of width $2 R_{\rm sat, projected} V_{\rm vir,sat}$, accounting for the spread of $j_{\rm sat}$ due to its rotation, and deposit the satellite's gas in the central's gas annuli of corresponding $j$ (see Fig.~\ref{fig:cartoon}e).  Each of these annuli are then subject to a merger starburst (Section \ref{ssec:starburst}) and \emph{direct} feeding of the black hole (equation \ref{eq:BHmajor}).  If the satellite's orbit is retrograde as seen from the central's gas disc, we consider twice the gas mass deposited in an annulus, less that consumed in the starburst, to be unstable (fed into the instability prescription above).

To be consistent with \textsc{sage}, and for simplicity, we deposit merging satellites' stars in the merger-driven bulge of the central.  We find that half of all minor mergers are retrograde relative to the central's disc.  Therefore, angular momentum for stars in minor mergers should be conserved on average.  The direct feeding of the black hole from the gas throughout the disc also removes some of the angular momentum we artificially introduce from retrograde satellites.  Our consideration of angular momentum during mergers is a first-order approximation only, and clearly has room for improvement.

\subsubsection{Merger starbursts}
\label{ssec:starburst}
Whenever gas originating from two galaxies collides in an annulus, a merger starburst occurs.  To determine how much gas is consumed in the starburst, we again maintain the prescription in \textsc{sage} \citep[from][]{somerville01} but apply it to the annuli individually:
\begin{multline}
\Delta m_{i, \rm burst} = - \beta_{\rm burst} (m_{i, \rm sat} + m_{i, \rm cen}) \\ \left[ \mathrm{min} \left(\frac{m_{i, \rm sat}}{m_{i, \rm cen}}, \frac{m_{i, \rm cen}}{m_{i, \rm sat}} \right) \right]^{\alpha_{\rm burst}}~,
\end{multline}
with $\alpha_{\rm burst} = 0.7$ and $\beta_{\rm burst} = 0.56$ \citep[see][]{mihos94,mihos96,cox04}.  All stars generated by merger starbursts are added to the merger-driven bulge.


\subsection{Black holes and active galactic nuclei}
\label{ssec:agn}

We maintain a two-mode model for black-hole growth and AGN feedback, with a `radio mode' and `quasar mode', and refer the reader to section 9 of \citet{sage} for a complete description.  The radio mode is implemented identically to \textsc{sage}.  Here, we slightly modify the quasar feedback prescription.  In \textsc{sage}, quasar feedback either heated all the cold gas in a galaxy or none of it.  With \textsc{Dark Sage}, we check if the energy is sufficient to heat the gas of each annulus, beginning from the centre and working outward until all the energy is used.  If there is still energy left to eject the hot gas, this will occur as well.


%% file: Sec5.tex
\section{Galaxy structure}
\label{sec:profiles}
The results we first present are the surface density profiles of Milky Way-like galaxies from \textsc{Dark Sage} at $z=0$.  Comparing these profiles to well-studied observed galaxies serves as a test for whether the model can produce galaxies with realistic spatial structure.  This is an important check of the model methodology.  More explicitly, this legitimises using specific angular momentum as an independent variable of galaxy evolution.  In order to make precise predictions for the net spin of galaxies and how this scales with other properties, one must integrate over the galaxies' angular-momentum structure.

\begin{figure}
	\centering
	\includegraphics[width=0.95\textwidth]{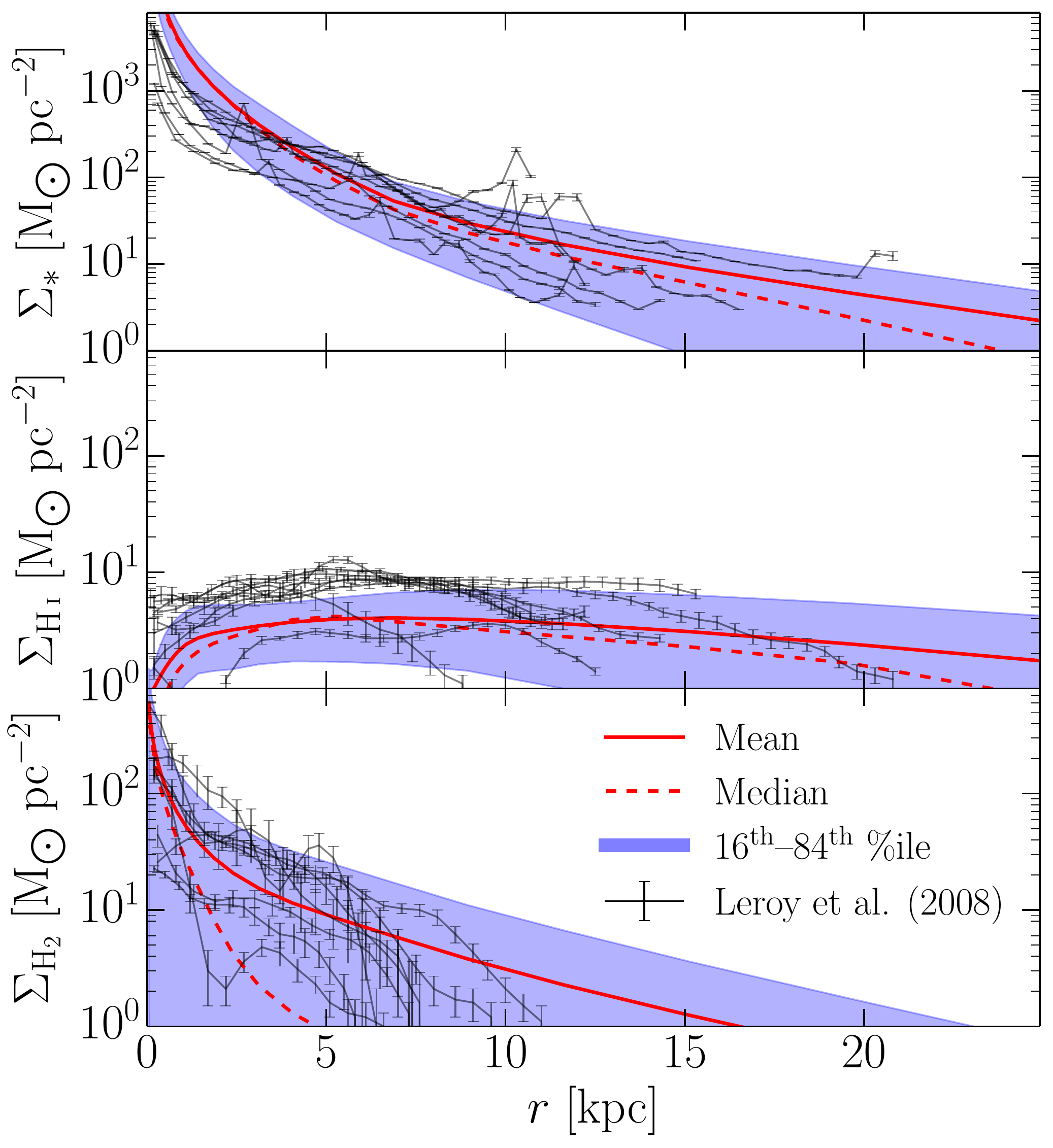}
	\caption{Face-on surface density profiles for galaxies produced by \textsc{Dark Sage} at $z=0$.  The three panels display the stellar, H\,\textsc{i}, and H$_2$ profiles, from top to bottom.  The gas profiles are built solely from that in discs, while the stellar profiles include the bulge mass contribution (note that $\Sigma_*$ refers only to the disc everywhere else in this paper).  We include galaxies with $175 \leq V_{\rm vir} / {\rm km\, s^{-1}} \leq 235$, $m_* > 10^{10}h^{-1}\,\mathrm{M}_{\odot}$, $m_{\rm cold} > 10^{9.2}h^{-1}\,\mathrm{M}_{\odot}$, and bulge-to-total ratios $<0.5$.  The mean, median, and 16$^{\rm th}$--84$^{\rm th}$ percentile range of the profiles of the \textsc{Dark Sage} galaxies within these cuts are presented.  In support, profiles of observed galaxies with similar properties are shown with errors \citep{leroy08}, specifically NGC 628, 3184, 3351, 3521, 3627, 5055, 5194, and 6946.}
	\label{fig:profiles}
\end{figure}

Because annuli have edges of fixed specific angular momentum, the corresponding disc radius changes from galaxy to galaxy, based on the given galaxy's rotation curve.  First, we interpolate the radial surface density profiles of the \textsc{Dark Sage} galaxies for each species of stars, H\,\textsc{i}, and H$_2$ onto a common radial grid. Then we plot these in Fig.~\ref{fig:profiles}, showing the 16$^{\rm th}$--84$^{\rm th}$ percentile range across the bins, as well as the mean and median.  With cuts to virial velocity, morphology, stellar and gas mass, these are compared against analogue, observed galaxies from \citet{leroy08}.  To the stellar profiles, we add the contribution from the bulge of each galaxy by projecting their assumed profiles (see Appendix \ref{app:rotcurves}) to two dimensions.

The predicted surface density profiles of \textsc{Dark Sage} galaxies are broadly in agreement with observations, at a level just as good, if not better, than previous semi-analytic attempts to build disc structure \citep[e.g.][]{fu10,fu13}.  Because the result of Fig.~\ref{fig:profiles} in no way affected the parameter values of the model, we can be relatively confident about the structure-dependent predictions the model makes.

We note there appears to be an excess in the central stellar surface density of our galaxies compared to observations (top panel of Fig.~\ref{fig:profiles}).  This contribution is from the disc and not a bulge component (but see Section \ref{ssec:discprofs}).  This feature has shown up consistently in models of disc evolution, especially those based on angular momentum \citep[e.g.][]{dalcanton97,bosch01,bullock01,stringer07}.  This excess could be explained by the model lacking consideration of radial dispersion support in the disc.  While the assumption that discs are entirely rotationally supported is accurate at larger radii, a comparable level of dispersion support should exist towards the centres of discs, which breaks down the simplicity of building a rotation curve and applying equation (\ref{eq:j}) to convert from $j$ to $r$.  As noted by \citet{bower06}, the assumption that discs shrink to be entirely rotationally supported can make them too small and almost completely self-gravitating, which is consistent with our results.  The mass contributions from the overdense regions of the discs are insufficient to notably affect their integrated properties. Our primary results of this paper, concerned with angular momentum, are hence not directly affected by this.

The extra central stellar content leads to an artificial rise in the central gas pressure (through equation \ref{eq:pressure}), which decreases the fraction of gas in an H\,\textsc{i} state.  As such, the \textsc{Dark Sage} galaxies also show a deficit in H\,\textsc{i} surface density in their centres (middle panel of Fig.~\ref{fig:profiles}).  This has a minimal effect on their integrated H\,\textsc{i} content.

\subsection{Discs and pseudobulges}
\label{ssec:discprofs}

\begin{figure}
	\centering
	\includegraphics[width=0.95\textwidth]{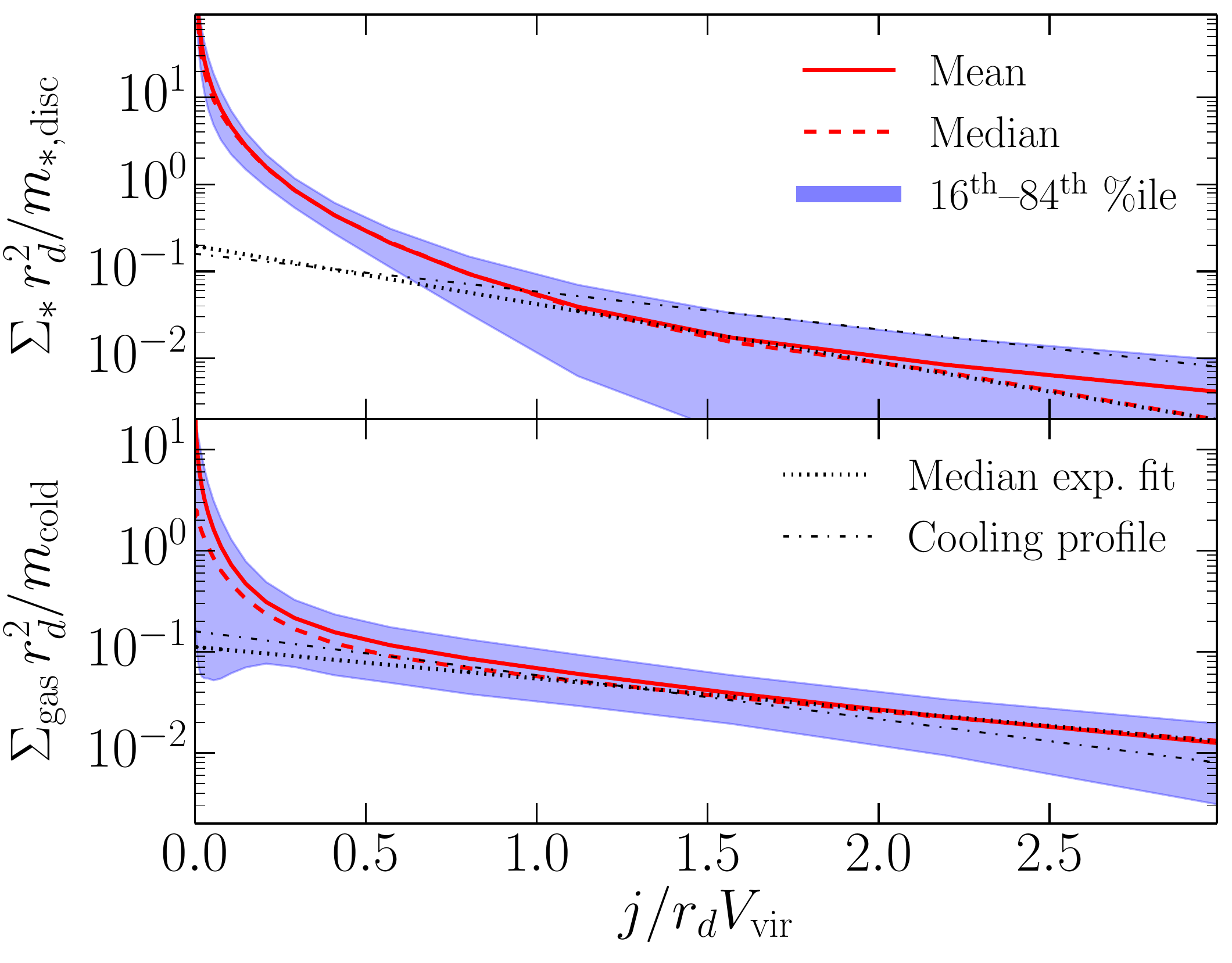}
	\caption{Face-on surface density profiles of \emph{discs} alone for stars (top) and cold gas (bottom) as a function of specific angular momentum.  Surface densities for each galaxy have been normalised by their mass and presumed scale radius (equation \ref{eq:r_d}), while $j$ has been normalised in each case by the same radius and the virial velocity of the halo.  We compare exponential profiles in each panel, fitted to the median curves over the range $j/r_d V_{\rm vir} > 0.9$. The dot-dashed line in the bottom panel represents equation (\ref{eq:cooling}). The same \textsc{Dark Sage} galaxy population is presented in Fig.~\ref{fig:profiles}. See text in Section \ref{sec:profiles} for a discussion.}
	\label{fig:disc_profiles}
\end{figure}

Although gas cools initially into a disc with an exponential dependence on $j$ in the model, come redshift zero, discs profiles are not exponential in the inner regions.  This is due to a culmination of processes: full rotation curve modelling leads to gas (and stars) settling into a more complex structure, various star formation channels convert gas to stars in different regions at different rates, feedback from star formation affects each annulus differently, instabilities drive material inward, etc.  We demonstrate this by plotting the surface density profiles of discs alone for stars and all cold gas as a function of $j$ in Fig.~\ref{fig:disc_profiles}.  The profiles for each galaxy have been normalised, such that they can be fairly compared with one another, and such that one would expect an exponential profile in the case that equation (\ref{eq:cooling}) were the only expression driving disc structure (which, obviously, is not the case).  An equivalent plot with $r/r_d$ on the $x$-axis looks nearly identical.  We have fitted exponentials to the medians of the outer parts of the profiles and over-plotted equation (\ref{eq:cooling}).  Because stars form primarily from low-$j$ gas (see Section \ref{ssec:sf}), it is unsurprising that these exponentials are consistent for gas but inconsistent for stars.

Few galaxies have been studied in the local Universe in sufficient detail to directly measure $\Sigma_{\rm gas}(r)$.  While many spirals have had their H\,\textsc{i} structure observed, H$_2$ data are required as well for a complete picture.  \citet{bigiel12} derived cumulative surface density profiles for H\,\textsc{i} and H$_2$ of 33 local spirals, promoting a `universal' profile from their results.  The normalised profiles of their galaxies (see their figs.~2 and 3) are consistent with an exponential gas disc with a central cusp.  Our results are, therefore, at least qualitatively consistent with observations, but more data are required to test this thoroughly.

Observationally, cusps at the centres of otherwise-exponential stellar surface density profiles are often associated with a bulge component.  In our model, while we have accounted for a purely pressure-supported instability-driven bulge, we have no explicit consideration of pseudobulges. Characteristically, pseudobulges have a steeper density profile than an exponential disc (with S\'{e}rsic indices $\lesssim$2) but are rotationally supported like discs \citep{kormendy93,kormendy04,fisher08}.  These can form through disc instabilities driving material towards the centre of galaxies and heating them, resulting in a rising thickness at lower radii \citep[see][]{toomre66,kormendy04}.  The cusp seen at low $j$ in our disc profiles matches this description, suggesting galaxies in \textsc{Dark Sage} naturally form pseudobulges.

\begin{figure}
	\centering
	\includegraphics[width=0.95\textwidth]{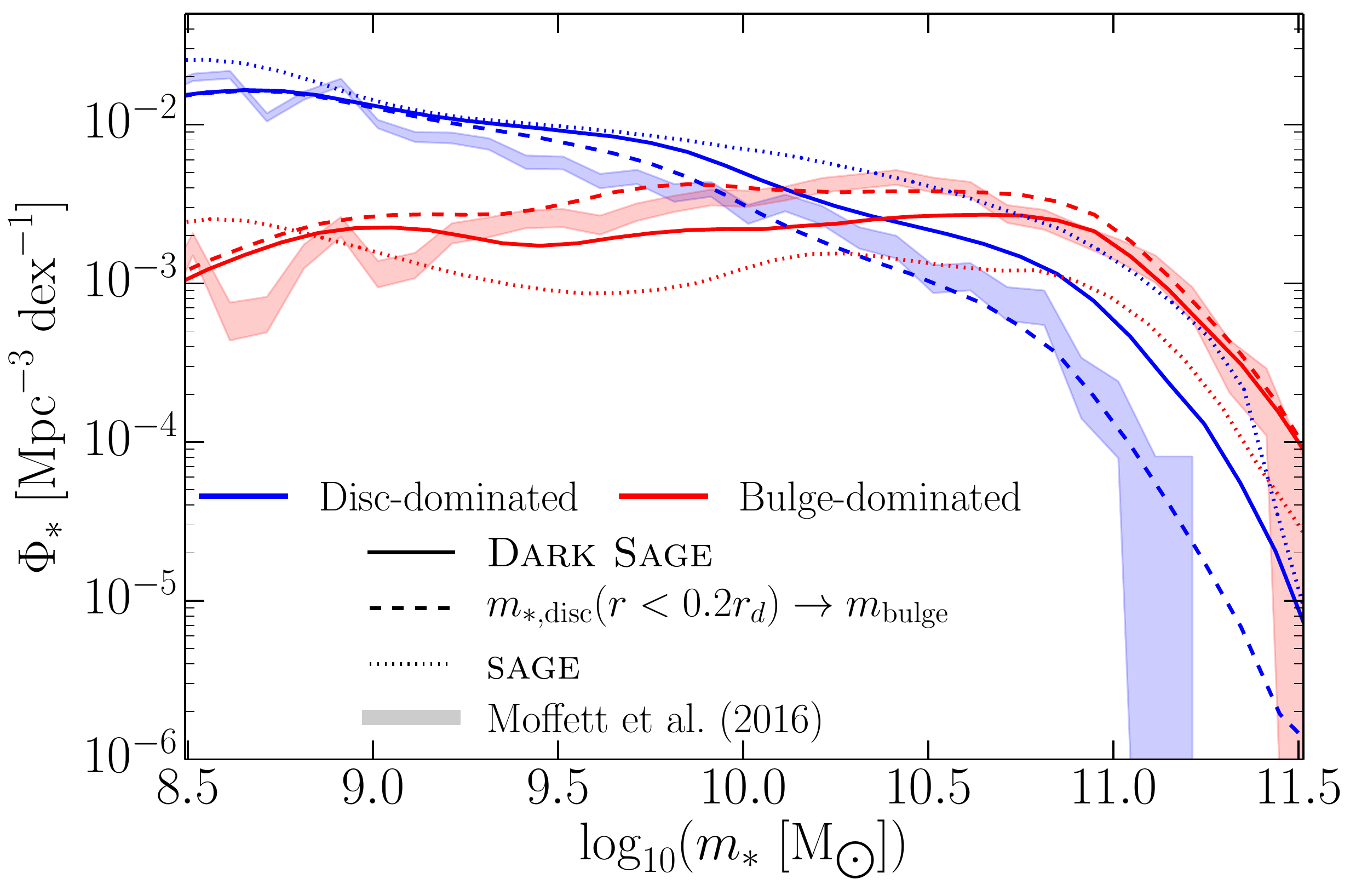}
	\caption{Stellar mass functions considering bulge-dominated (bulge-to-total ratio $>0.5$) and disc-dominated ($\leq 0.5$) systems separately.  Solid curves are the direct output of \textsc{Dark Sage} at $z=0$, whereas the dashed curves assume that stellar disc mass inside $0.2 r_d$ counts as pseudobulge mass for systems that have developed an instability-driven bulge.  Compared is the result from the public version of \textsc{sage} and observations from \citet[][based on Hubble type]{moffett16}.}
	\label{fig:smf_morph}
\end{figure}

\citet{fisher08} have empirically shown that pseudobulges, on average, have half-mass radii approximately equal to $0.2 r_d$.  If we consider, in post-processing, for galaxies that have formed an instability-driven bulge, all the stellar disc mass within $0.2 r_d$ to be a part of a bulge instead of the disc, we naturally find different bulge-to-total ratios for our galaxies.  To demonstrate this, we have compared the results of \textsc{Dark Sage} with and without this redistribution of mass to observations of the bulge-dominated and disc-dominated stellar mass functions \citep{moffett16}.  This is presented in Fig.~\ref{fig:smf_morph}.  We find overall better agreement with the data with this simplistic pseudobulge consideration (dashed curves) than without (solid curves).  In both cases, \textsc{Dark Sage} shows improvement from the public version of \textsc{sage} (dotted curves).  We note, however, that for the model galaxies, disc-dominated means a bulge-to-total ratio less than 0.5, whereas the terms `disc-dominated' and `bulge-dominated' in \citet{moffett16} are based on visual Hubble classifications.  As such, this comparison to observations should be considered approximate.

%% file: Sec6.tex
\section{Mass--spin relation of spirals}
\label{sec:mj}

Several observational studies of galaxies in the local Universe have suggested there is a strong correlation between the stellar mass of a galaxy and its specific angular momentum \citep[e.g.][]{fall83,rf12,og14}.  These mass--spin sequences evidently vary for galaxies of different morphology.

\textsc{Dark Sage} evolves the angular-momentum structure of discs, but makes no direct predictions for the angular momentum of spheroids.  In the interest of comparing against larger datasets, we focus on results concerning the stellar specific angular momentum of spiral galaxies.  We intend to address the total baryonic specific angular momentum of \textsc{Dark Sage} galaxies in future work.

\subsection{Stellar discs in the local Universe}
\label{ssec:discs}

\begin{figure}
	\centering
	\includegraphics[width=0.95\textwidth]{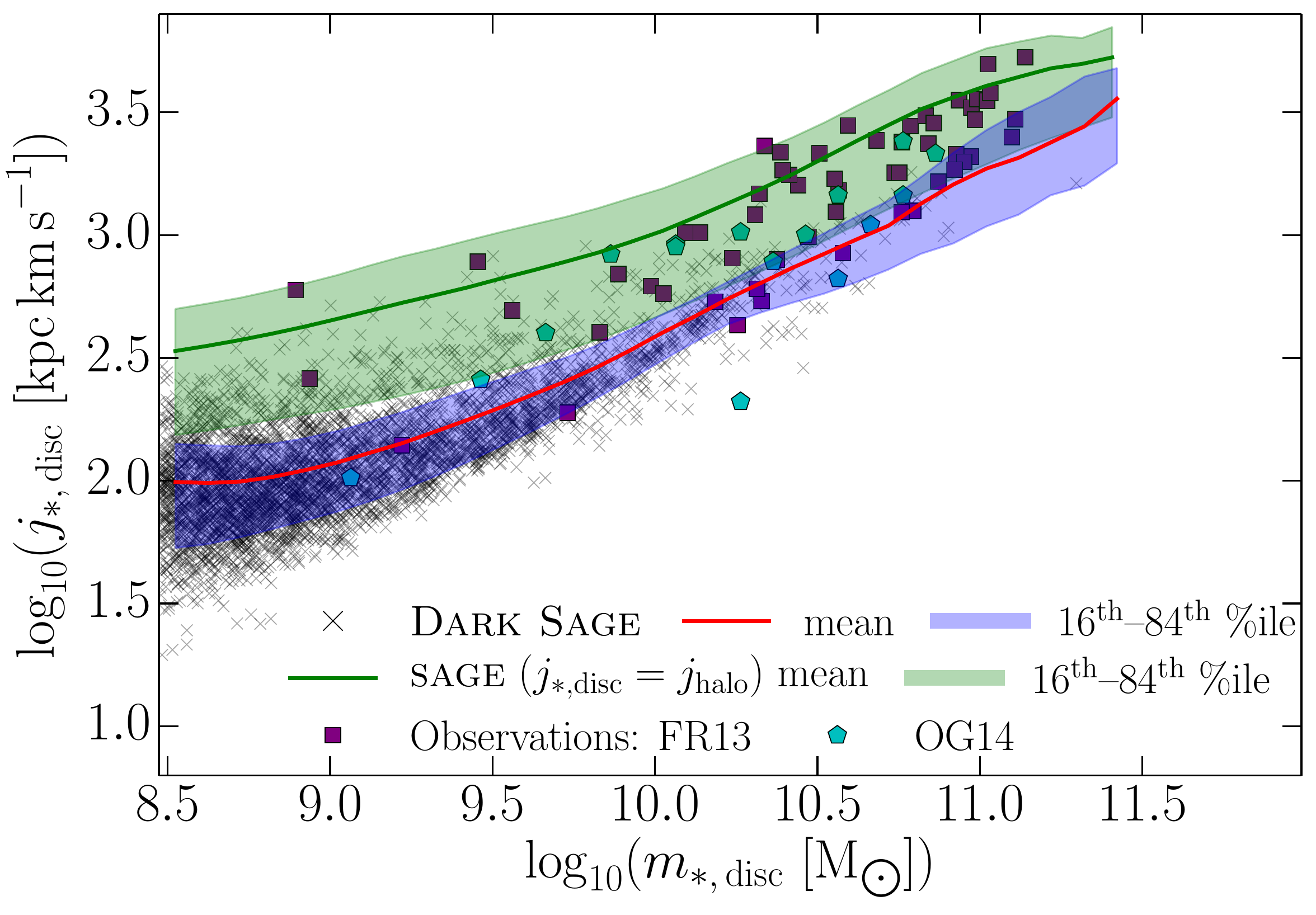}
	\caption{Integrated stellar \emph{disc} mass versus net specific angular momentum of stellar \emph{discs} from \textsc{Dark Sage} galaxies with bulge-to-total ratios $< 0.3$.  4000 randomly sampled galaxies are shown by the black crosses in the axes, with the mean and inner 68\% of galaxies in 0.1-dex bins also presented.  Compared is the equivalent result from \textsc{sage}, which assumes the specific angular momentum of the disc and halo are the same.  Supporting observational data from spiral galaxies are presented from \citet{fr13} and \citet{og14}.  Their uncertainties in $j_{*,\, {\rm disc}}$ are approximately the size of the data points, while the mass uncertainties are dominated by systematics at the level of 0.5 dex \citep[see section 3 of][]{rf12}.}
	\label{fig:jm}
\end{figure}

With Fig.~\ref{fig:jm}, we display the predicted stellar mass--spin relation at $z=0$ for discs of spiral galaxies from \textsc{Dark Sage}, i.e.~showing integrated mass, $m_{*,\,{\rm disc}}$, against net specific angular momentum,
\begin{equation}
j_{*,\,{\rm disc}} \equiv \frac{\sum_{i=1}^{30} m_{*i}\, \bar{j}_i}{\sum_{i=1}^{30} m_{*i}} = \frac{\sum_{i=1}^{30} m_{*i}\, (j_i + j_{i-1})}{2\, m_{*,\,{\rm disc}}}~.
\end{equation}
We use galaxies with bulge-to-total ratios less than 0.3 as `spiral galaxies'.  Compared in Fig.~\ref{fig:jm} are the equivalent results from the public version of \textsc{sage} \citep{sage} and observations of spiral galaxies \citep{fr13,og14}.
 
In \textsc{sage}, $j$ is not an evolved quantity; instead, it is assumed that $j_{*,{\rm disc}} = j_{\rm halo}$.  While $j_{\rm halo}$ is set entirely by the simulation and not the semi-analytic model, the corresponding $m_{*,{\rm disc}}$ comes from the galaxy evolution physics.  This is indirectly and loosely constrained by the stellar mass function.  The lack of two-dimensional freedom limits \textsc{sage}'s ability to examine the cause of the mass--spin relation of discs.  With its complete evolution of angular-momentum structure, \textsc{Dark Sage} is better equipped.

That said, both models overlap the observational data, where the data seem to lie in between the two models (Fig.~\ref{fig:jm}).  It is unsurprising that the \textsc{Dark Sage} sequence has systematically lower $j$ than \textsc{sage}; while gaseous discs carry the same spin as the halo upon birth, subsequent cooling episodes, gas precession, and mergers in \textsc{Dark Sage} all lead to a reduction in the specific angular momentum of the disc.  In addition, stars are preferentially born from low-$j$ gas.  The \textsc{Dark Sage} sequence is also notably tighter than \textsc{sage}.

The observational data we have compared against have been measured with two techniques.  Many of the \citet{fr13} data calculated $j_{*,\rm disc}$ analytically by assuming stellar discs followed exponential surface density profiles with a constant rotation velocity.  These were originally published in \citet{rf12}, but the masses were updated in \citet{fr13}, using a variable mass-to-light ratio.  The \citet{og14} specific angular momenta were calculated by properly integrating the resolved angular momentum profiles, but their masses assumed a constant mass-to-light ratio.  These authors find the uncertainty introduced by calculating $j_{*,\rm disc}$ with the method of \citet{rf12} is correlated with the value of $j_{*,\rm disc}$.  Six galaxies exist in both samples.

Earlier in this project, we hypothesised that the instability prescription could drive the mass--spin relation of discs.  This is because the prescription, in essence, places an upper limit on the amount of mass allowed in an annulus.  As instabilities drive most of the unstable mass inward, low-angular-momentum mass from the disc is transferred to the bulge.  This then approximately translates into a lower limit on the net specific angular momentum in a stellar disc of a given integrated mass.

\begin{figure}
	\centering
	\includegraphics[width=0.95\textwidth]{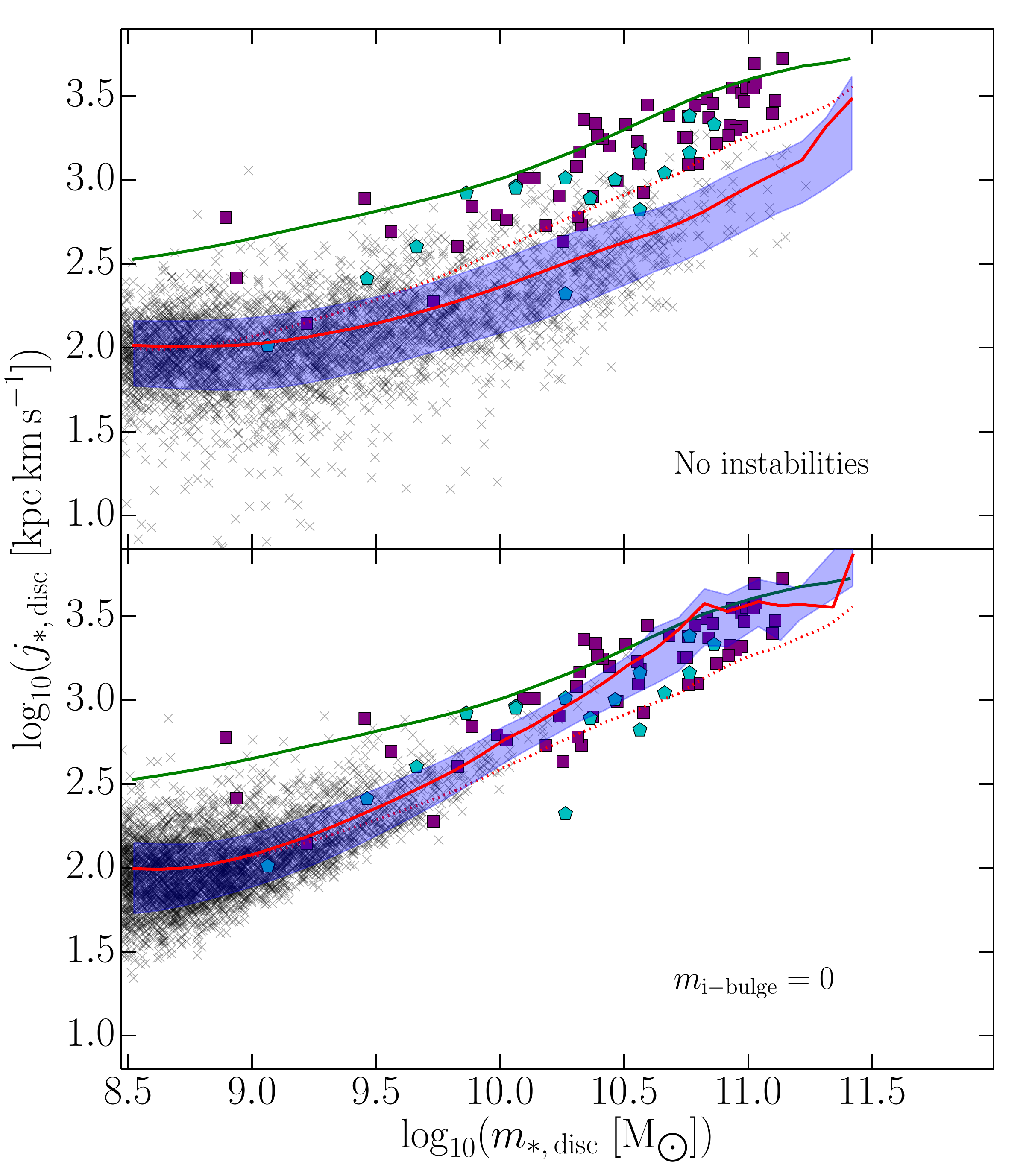}
	\caption{\emph{Top panel}: As for Fig.~\ref{fig:jm} but now excluding instabilities from \textsc{Dark Sage} entirely. \emph{Bottom panel}: A subsample of galaxies from the fiducial run of \textsc{Dark Sage}, excluding all systems with non-zero instability-driven bulge mass.  The dotted line is the mean of \textsc{Dark Sage} from Fig.~\ref{fig:jm}. See text in Section \ref{ssec:discs}.}
	\label{fig:jm_extra}
\end{figure}

The top panel of Fig.~\ref{fig:jm_extra} presents the mass--spin relation for \textsc{Dark Sage} stellar discs if all consideration of disc instabilities is removed from the model (without changing parameter values).  With instabilities switched off, the low-$j$ part of the axes becomes occupied.  In addition, the average $j_{*,\,{\rm disc}}$ is lowered (e.g.~at $m_{*,\rm disc} = 10^{11}\,\mathrm{M}_{\odot}$, it is lower by $\sim$0.4 dex).  The scatter (standard deviation) also increases from $\sim$0.19 to $\sim$0.22 dex.  There still exists a trend between mass and spin, but without instabilities, the model would simply not be able to match the data.

Although the above suggests instabilities drive the mass--spin relation of discs, some galaxies may have had stable discs since their formation.  We can identify those galaxies in \textsc{Dark Sage} as having no instability-driven bulge (if these galaxies have had major mergers, their discs will have been stable since the last major merger).  We can consider this group of galaxies somewhat as a control sample to test whether stable galaxies still obey the observed mass--spin relation.  These galaxies are shown in the bottom panel of Fig.~\ref{fig:jm_extra} and match the observational data almost perfectly.  Indeed, if we use the halo identification numbers of these galaxies and rerun the model without instabilities, these systems still match the observations.  Therefore, the galaxy evolution physics considered in \textsc{Dark Sage} (disregarding instabilities) naturally explains the mass--spin relation of spiral galaxies with stable discs.  Toomre instabilities then act to regulate the specific angular momentum in the discs of the remaining galaxies to naturally bring them in line with observations.

\begin{figure}
	\centering
	\includegraphics[width=0.95\textwidth]{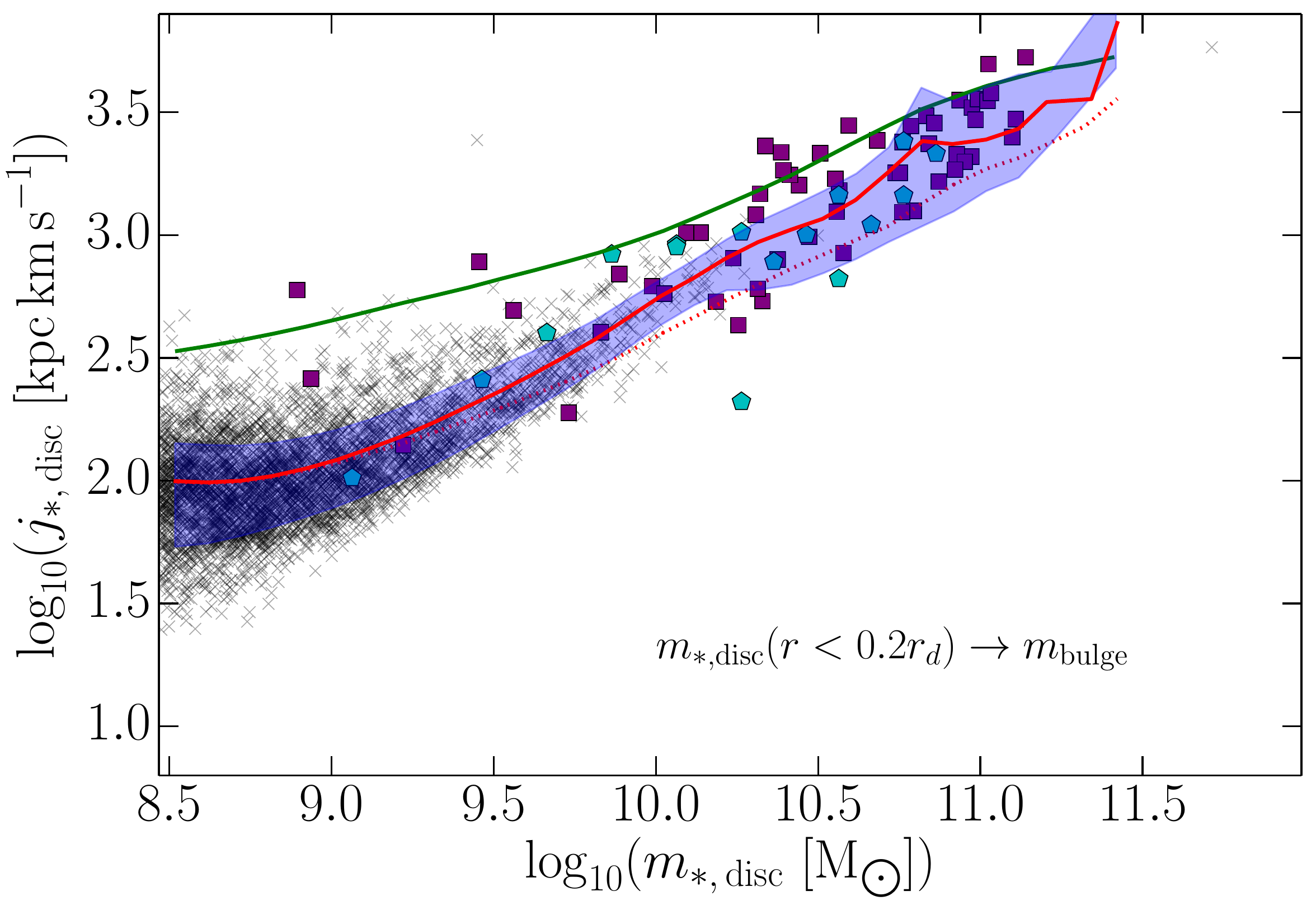}
	\caption{As for Fig.~\ref{fig:jm}, but any stellar disc material inside $0.2 r_d$ for \textsc{Dark Sage} galaxies with instability-driven bulges is associated with the bulge instead.  The dotted line compares the mean of \textsc{Dark Sage} without this adjustment (from Fig.~\ref{fig:jm}).}
	\label{fig:jm_inner}
\end{figure}

 As discussed in Section \ref{ssec:discprofs}, the inner disc material of \textsc{Dark Sage} galaxies may actually represent a pseudobulge.  In principle then, the \textsc{Dark Sage} discs should have lower masses and higher specific angular momenta than that presented in Fig.~\ref{fig:jm}.  Once again, to approximate this to first order, we shift all stellar mass within $0.2 r_d$ to the bulge in post-processing for the galaxies with non-zero instability-driven bulge masses.  We then resample the spiral galaxies and plot the mass and specific angular momentum of their stellar discs in Fig.~\ref{fig:jm_inner}.  With this adjustment, we now find \textsc{Dark Sage} spiral galaxies in general to be in remarkable agreement with observations.
 
Pseudobulges are thought to form through disc instabilities, so, naturally, galaxies without an instability-driven bulge should be without a pseudobulge.  Hence the results of the mass--spin relation after adjusting for the pseudobulges of general \textsc{Dark Sage} spiral galaxies (Fig.~\ref{fig:jm_inner}) agree quite well with the spiral galaxies without instability-driven bulges (bottom panel of Fig.~\ref{fig:jm_extra}).
 
\subsection{Evolution of stellar discs}

\begin{figure}
	\centering
	\includegraphics[width=0.95\textwidth]{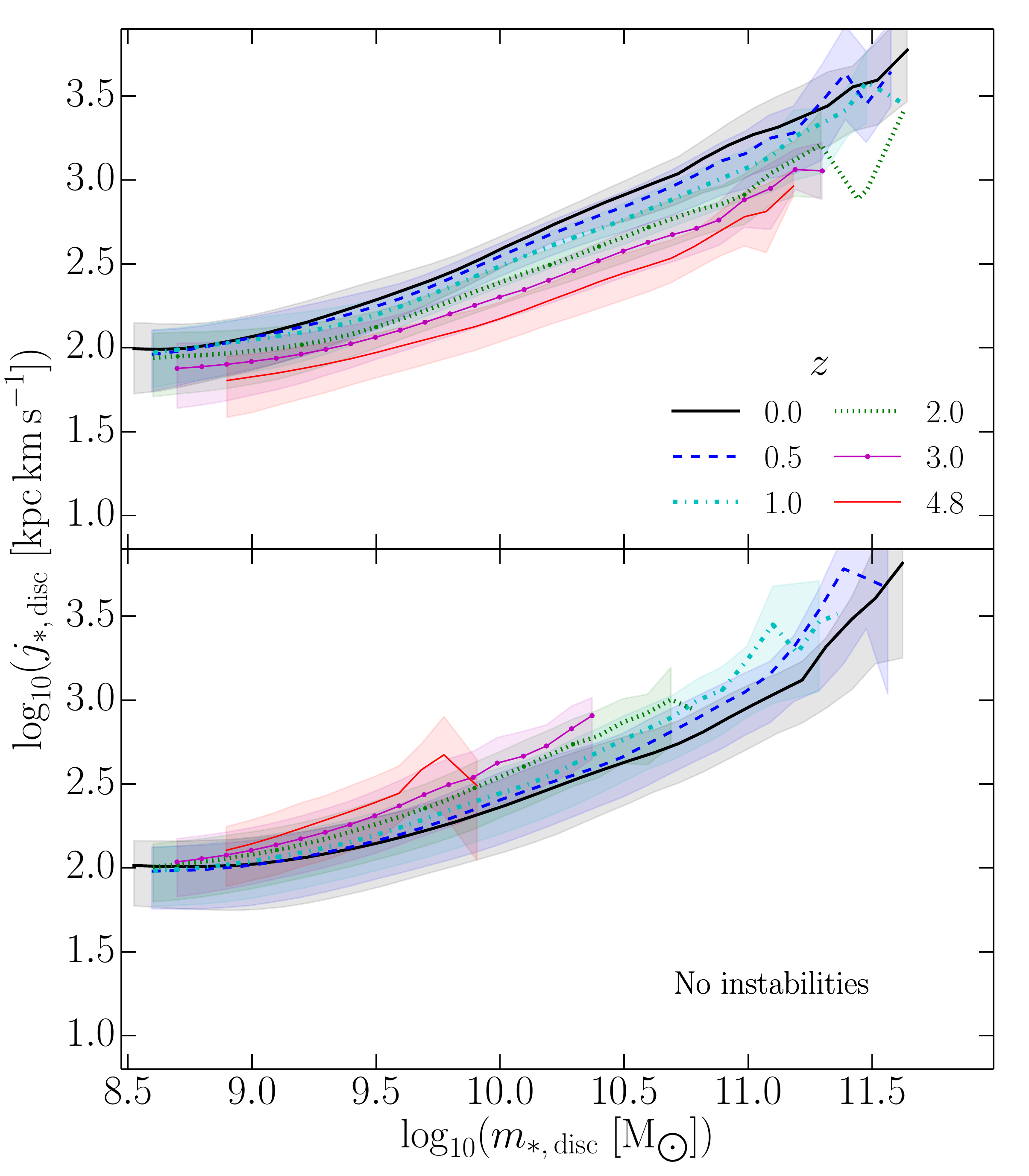}
	\caption{Evolution of the mass--spin sequence for stellar discs of \textsc{Dark Sage} galaxies with bulge-to-total ratios $<0.3$.  The shaded region around each curve covers the 16$^{\rm th}$--84$^{\rm th}$ percentile range for that redshift.  \emph{Top panel}: Fiducial model.  \emph{Bottom panel}: Instability prescription excluded from the model.}
	\label{fig:jm_evo}
\end{figure}

We now turn our attention to the redshift evolution of the mass--spin relation for stellar discs of spiral galaxies.  In the top panel of Fig.~\ref{fig:jm_evo}, we present the sequence from \textsc{Dark Sage} at select redshifts, where we have selected galaxies based on their morphology at that redshift.  We find there is a trend for stellar discs of fixed mass to have lower specific angular momentum at earlier epochs.  The trend is relatively weak however, where the mean of the sequence only increases by $\sim$0.4 dex from $z=4.8$ to 0 at the high-mass end.  This compares to a standard deviation about each sequence of $\sim$0.15 dex.  Qualitatively, the same trend is seen in the EAGLE hydrodynamic simulations (Lagos et al., in preparation).

In the bottom panel of Fig.~\ref{fig:jm_evo}, we present the same mass--spin sequence evolution, but now without instabilities in the model.  Not only does this alter the sequence for the local Universe, but the dependence with time is \emph{opposite} to the full model.  Without the instability channel of star formation, stellar discs take longer to build up their mass.  Also, without the ability to regulate where in the disc this stellar mass ends up, these discs are typically low-mass with high specific angular momentum at higher redshift.  As time progresses, the low-$j$ stars, that otherwise would have been transferred to the bulge, loiter in the disc.

\begin{figure}
	\centering
	\includegraphics[width=0.95\textwidth]{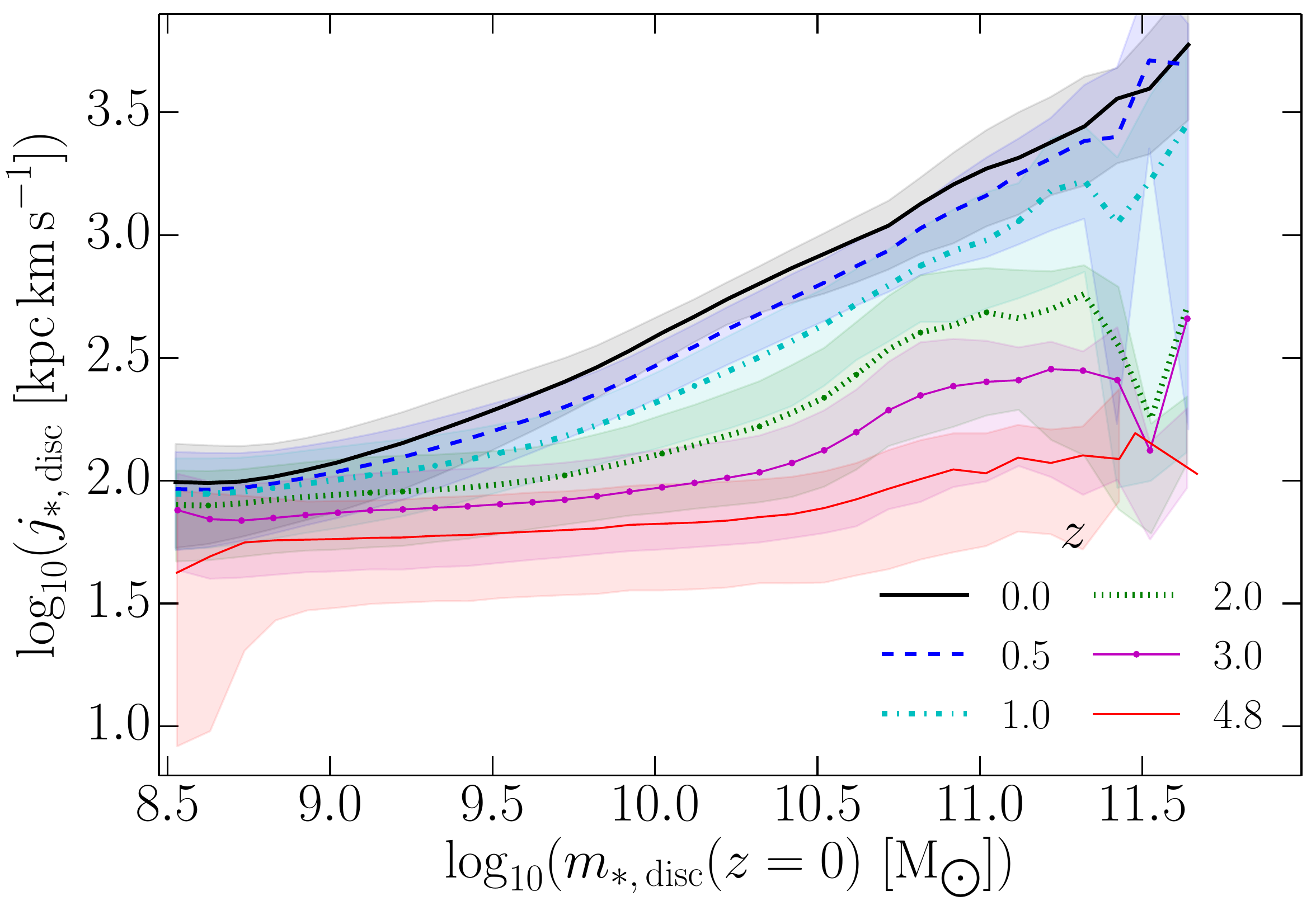}
	\caption{Evolution of the specific angular momentum of stellar discs of the main progenitors of $z=0$ spiral galaxies in \textsc{Dark Sage}, measured as a function of their mass at $z=0$.  Plotting style follows Fig.~\ref{fig:jm_evo}.}
	\label{fig:jm_evo_z0}
\end{figure}

Even without instabilities, $j_{*,\,{\rm disc}}$ of a given galaxy does not typically decrease with time.  To illustrate this point, we show the evolution of $j_{*,\,{\rm disc}}$ for the most-massive progenitors of $z=0$ spirals in \textsc{Dark Sage} in Fig.~\ref{fig:jm_evo_z0}, given as a function of their final mass.  These results include instabilities in the model, but the same qualitative result is found without them.  Here we see that stellar discs, on average, clearly begin with lower specific angular momentum than they end up with.  

As shown by Fig.~\ref{fig:jm_evo_z0}, as we approach higher redshift, the mean spin of the local-spiral progenitors begins to lose its dependence on final mass.  By design, gas discs are initially set to have the same spin as their halo.  The spin of haloes follows a fairly strict log-normal distribution which is only weakly correlated with mass \citep{barnes87,bullock01,knebe08}.  Further, because the gas discs have the same spin at birth on average, the first episode of star formation in these discs will mean the same is true for stellar discs.  As such, the initial spin of a stellar disc should not depend on its final mass.  Note that larger objects in the local Universe tend to have formed (resolved structures in the simulation) at earlier epochs.  One thus has to go to relatively higher redshift to find the progenitors of local massive spirals at a state of equivalently low spin.

\subsection{Total stellar content} 
\label{ssec:all}
 
Given a halo with a known mass profile and spin parameter (equation \ref{eq:lambda}), one can calculate the halo's specific angular momentum.  Based on this idea, \citet{rf12} proposed a toy model for producing the mass--spin relation of galaxies, considering their total stellar content.  Those authors approximated the mass profiles of haloes as NFW \citep*{nfw96,nfw97} profiles with a fixed concentration, $c = 9.7$ (defined in Appendix \ref{app:rotcurves}).  Following this same process, but using the cosmology of the Millennium simulation and adjusting for the difference in definition of virial mass, which are implicitly assumed, this gives
\begin{equation}
\frac{j_{\rm halo}}{\rm kpc\, km\, s^{-1}} \simeq 5.73 \times 10^4\ \lambda \left( \frac{M_{\rm vir}}{10^{12}\,\mathrm{M}_{\odot}} \right)^{2/3}
\label{eq:jhalo}
\end{equation}
\citep[cf.~equation 14 of][]{rf12}.  The authors then related these quantities to their stellar equivalents assuming the stellar specific angular momentum of a galaxy to be some fixed fraction of the halo,\footnote{In \citet{rf12}, $f_j$ is discussed as the fraction of specific angular momentum lost by stars to other parts of the halo over time.  Given that equation (\ref{eq:jhalo}) assumes $z=0$, in addition to the fact that masses and spins of haloes are dynamic quantities, their description is not to be taken literally.}
\begin{equation}
f_j \equiv j_* / j_{\rm halo}~,
\end{equation}
and using the stellar--halo mass relation of \citet[][their equation 3]{dutton10}.  Under our implicit assumptions, this gives
\begin{equation}
f_*(m_*) \equiv \frac{m_*}{f_b\,M_{\rm vir}} \simeq 0.28 \sqrt{\frac{m_* / M_0}{1 + m_*/M_0}}~,
\end{equation}
where $M_0 = 10^{10.4} h^{-2}\, \mathrm{M}_{\odot}$.  Combining all of this leads to an expression for estimating $j_*$ from $\lambda$ and $m_*$ alone:
\begin{equation}
\frac{j_*}{\rm kpc\, km\, s^{-1}} \simeq 4.02 \times 10^4\ \frac{\lambda\, f_j}{f_*^{2/3}(m_*)} \left( \frac{m_*}{10^{11}\,\mathrm{M}_{\odot}} \right)^{2/3}~.
\label{eq:jstar_rf12}
\end{equation}
 
\begin{figure}
	\centering
	\includegraphics[width=0.95\textwidth]{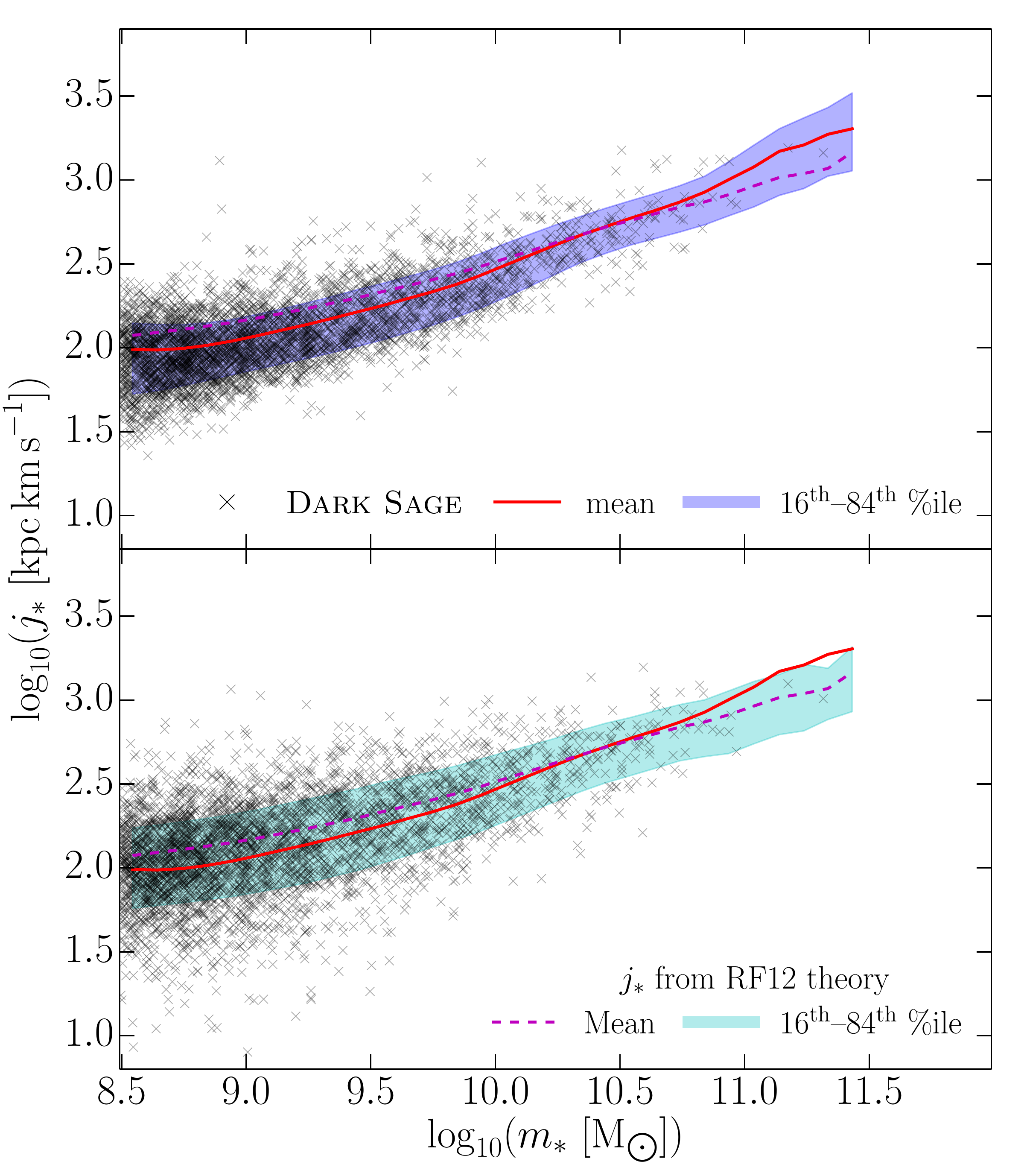}
	\caption{\emph{Top panel}: As for Fig.~\ref{fig:jm} but now calculating mass and specific angular momentum considering \emph{all stars} in \textsc{Dark Sage} spiral galaxies.  \emph{Bottom panel}: Reproducing the top figure using the same stellar masses, but calculating $j_*$ from the theory of \citet{rf12}, our equation (\ref{eq:jstar_rf12}), assuming $f_j = 0.16$.  We include the mean trends for values of $j_*$ in each case on both plots for comparison.}
	\label{fig:jm_all}
\end{figure}

We compare this theory to the direct predictions of \textsc{Dark Sage}.  In Fig.~\ref{fig:jm_all}, we again plot the mass--spin relation for spiral galaxies, but now include stars both in the disc and bulge.  We remind the reader that \textsc{Dark Sage} currently makes no direct predictions for the magnitude of angular momentum present in a bulge.  As such, we assume bulge stars to carry zero angular momentum.\footnote{With the exception of potential pseudobulges, which are incorporated in the disc profiles (Section \ref{ssec:discprofs}).}  That is,
\begin{equation}
j_* \equiv \frac{\sum_{i=1}^{30} m_{*i}\, (j_i + j_{i-1})}{2\, m_*}~.
\end{equation}
This is a reasonable approximation, as, based on our morphology cut, bulge stars should contribute very little angular momentum compared to those in the disc \citep[see section 2.3 of][]{og14}.

To produce values of $j_*$ from the toy model of \citet{rf12}, we take the same sample of \textsc{Dark Sage} galaxies, and plug their values of $\lambda$ and $m_*$ into equation (\ref{eq:jstar_rf12}).  In the bottom panel of Fig.~\ref{fig:jm_all}, we plot the derived mass--spin relation, assuming $f_j = 0.16$, a value we treated freely to find the best visual agreement.  The mean relation is similar to the true \textsc{Dark Sage} one, but the toy model does induce a larger amount of scatter (standard deviations of $\sim$0.18 dex versus $\sim$0.25 dex).  As a point of interest, had we instead presented \textsc{Dark Sage} results without the instability prescription here, the scatter would have been consistent.

\begin{figure}
	\centering
	\includegraphics[width=0.95\textwidth]{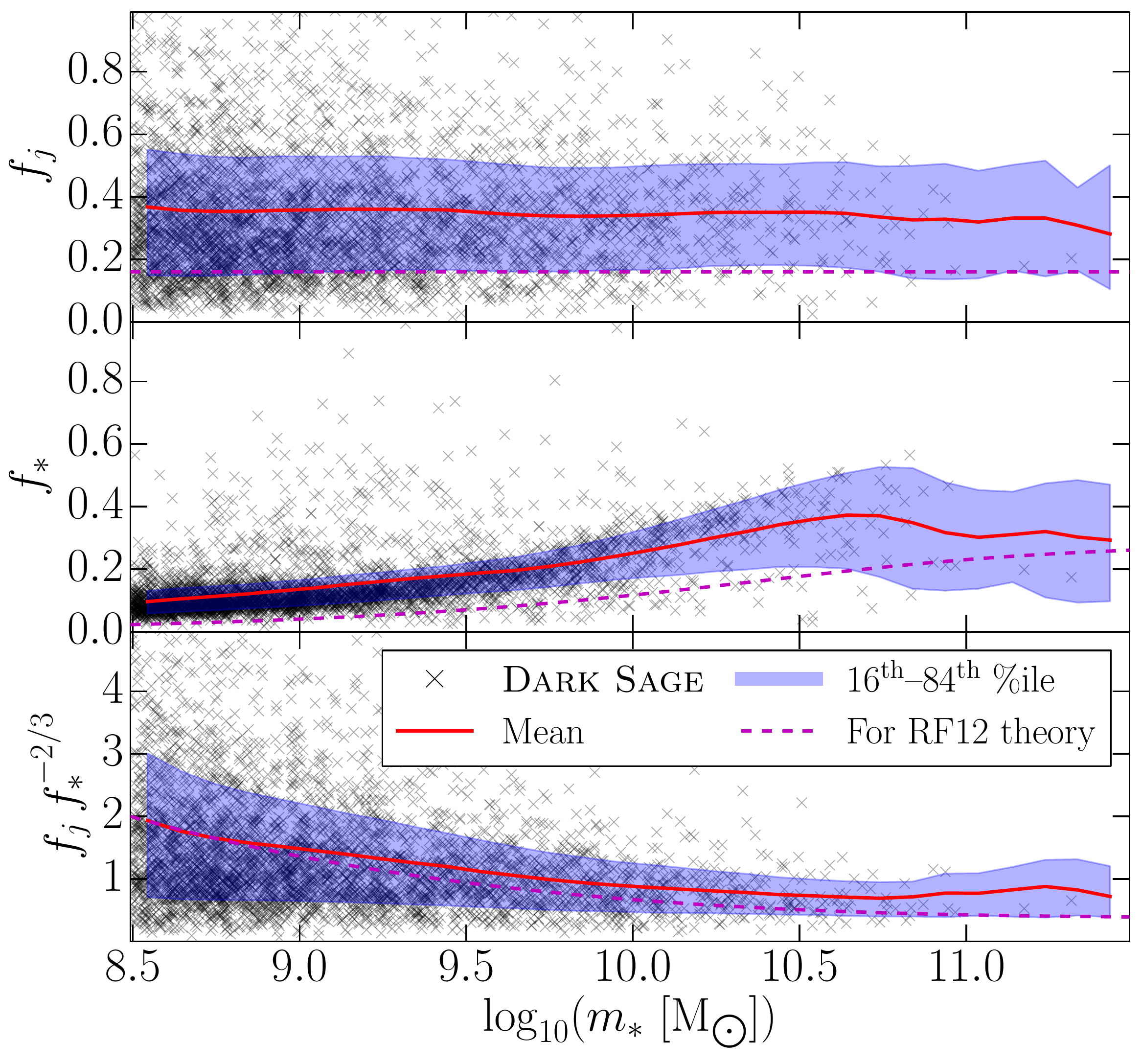}
	\caption{\emph{Top panel}: Ratio of specific angular momentum in stars to that of the halo for the same sample of \textsc{Dark Sage} spiral galaxies in Figs \ref{fig:jm} and \ref{fig:jm_all}.  \emph{Middle panel}: Ratio of stellar mass to baryonic mass in those same haloes.  \emph{Bottom panel}: Observationally constrainable combination of the other two quantities, relevant for equation (\ref{eq:jstar_rf12}).  In each of the three cases, the solid curve is the mean of the \textsc{Dark Sage} galaxies and the shaded region covers the inner 68\% of galaxies, while the dashed curve represents the values assumed for calculating the \citet{rf12} model predictions of $j_*$, as presented in Fig.~\ref{fig:jm_all}.}
	\label{fig:fj_fstar}
\end{figure}

Of course, the quantities $f_j$ and $f_*$ are also predicted directly by \textsc{Dark Sage}.  The first two panels of Fig.~\ref{fig:fj_fstar} present these values for the same galaxies as shown in Fig.~\ref{fig:jm_all}.  Both quantities, especially $f_j$, show a large amount of scatter.  For reference, the dashed lines also display the $f_j$ and $f_*$ values used above (0.16 and equation \ref{eq:jstar_rf12}, respectively).  Both of these lines are clearly below the actual average values of the galaxies for a given mass.  For example, we predict a mean value of $f_j \simeq 0.4$ for spiral galaxies.  This highlights the degeneracy in the value of $f_j$ and normalisation of $f_*$ when estimating $j_*$ for a galaxy with equation (\ref{eq:jstar_rf12}).  As discussed by \citet{rf12}, it is $f_j\,f_*^{-2/3}$ that is the quantity one can constrain the mean for with observations.  This quantity is also presented in Fig.~\ref{fig:fj_fstar}, and it too has a large amount of scatter from the \textsc{Dark Sage} spiral galaxies.  One should, therefore, heed caution when drawing conclusions about halo spin from stellar mass and angular momentum of individual galaxies.

%% file: Sec8.tex
\section{Outlook and conclusions}
\label{sec:outlook}

We have presented our new semi-analytic model of galaxy evolution, \textsc{Dark Sage}.  With this model, we have broken discs into annuli of fixed specific angular momentum, while evolving the angular momentum vectors of both the stellar and gas components of discs.  This level of detail in the angular-momentum evolution of galaxies is a new step for semi-analytic models, and allows processes such as star formation, supernova feedback, metal enrichment, ram-pressure stripping of cold gas in satellites, and disc instabilities to be performed locally.  Galaxies are therefore evolved as an explicit function of specific angular momentum, on local scales, self-consistently, in a cosmological framework.

In this paper, we have focussed on predictions of \textsc{Dark Sage} for two scientific areas of interest.  First, the model is capable of producing realistic surface density profiles for disc galaxies in the local Universe for each of stars, H\,\textsc{i}, and H$_2$, as presented in Section \ref{sec:profiles}.  Interestingly, the profiles of the discs themselves are \emph{not} exponential in the centres, but rather form a cusp, likely indicating the formation of a pseudobulge.  A first-order consideration of the mass of these pseudobulges brings the model in better agreement with observed morphological stellar mass functions.

Second, we have shown \textsc{Dark Sage} produces a clear sequence for the net specific angular momentum and mass of stellar discs in spiral galaxies, presented in Section \ref{sec:mj}.  This is in very good agreement with observations, especially after accounting for possible pseudobulges.  We found that a Toomre instability criterion plays an important role in reducing the scatter of this sequence and for improving its agreement with observations.  The minority of high-mass discs that suffer no instabilities since their last major merger naturally agree precisely with observations as well.  We also found the instability prescription determines the direction in which the mass--spin sequence evolves with redshift (left to right versus bottom to top).  The main progenitors of individual spiral galaxies at $z=0$ typically grow in specific angular momentum over their life-time, though, regardless of instabilities.  Finally, excluding angular momentum from a merger-driven bulge, our model predicts the ratio of the stellar specific angular momentum of a galaxy to its halo is 0.4 on average, with a standard deviation of 0.29.

\textsc{Dark Sage} is a model well suited for a number of different scientific investigations.  For example, we intend to focus on the metallicity gradients of nearby spiral galaxies in a future study.  Predictions for the model concerning H\,\textsc{i} scaling relations with respect to galaxy environment are to be presented in an upcoming paper as well.  In another work, we will focus on the different evolutionary tracks galaxies take in the \textsc{sage} and \textsc{Dark Sage} models by making object-to-object comparisons for the same $N$-body simulation haloes.

Plenty of room for further detail in models like \textsc{Dark Sage} remains.  For example, we only evolve the structure of discs in one dimension, thereby assuming axial symmetry.  The semi-analytic method does not lend itself well to detailed multi-dimensional information about galaxies, but some additions could be made.  Where we currently track a single angular-momentum vector for each gas and stellar disc, one could, in principle, track an angular-momentum vector for each \emph{annulus}, which would allow for the study of disc warps.  We also enforce all baryons in the disc to have positive specific angular momentum with respect to the net rotation axis, but hydrodynamic simulations have shown a fair portion should have retrograde motion \citep*[e.g.][]{chen03}.  Bins for negative values of $j$ could be included to account for this.  Furthermore, \textsc{Dark Sage} has no explicit consideration of bars, which could affect the internal disc structure of galaxies and how instabilities are dealt with.  The model could also include a more explicit consideration of radial migration.  Currently, radial migration happens naturally through binning discs by $j$ (Section \ref{ssec:reservoirs}) and resolving disc instabilities (Section \ref{ssec:instab}), but one may favour explicitly modelling diffusion, for example. 

While the physics of this model is founded on widely used theory, aspects of this model can, and should, be tested.  Specifically, the prescriptions for how gas settles into discs during cooling episodes and mergers could be compared with hydrodynamic simulations.  It has already been shown, for example, that the net specific angular momentum of dark matter and hot gas in haloes differs both in direction and magnitude (\citealt{bosch02,chen03}; \citealt*{bosch03}). The process of how this gas cools is vital to the evolution of angular-momentum structure in galaxies.  We are in the process of studying this, using the EAGLE simulations (Stevens et al.~in preparation).

%% file: Sec4.tex
\section{Model calibration}
\label{sec:constraints}

We have constrained the free parameters of \textsc{Dark Sage} by hand to match a set of observables, focussed on the local galaxy population.  Many of these constraints are identical to those used for \textsc{sage}.  We used the smaller `Mini Millennium' simulation, with a box length of $62.5 h^{-1}$\,Mpc (but otherwise matching the details in Section \ref{sec:sims}), for calibrating most of the parameters.  The plots presented in this appendix were preferably made with the larger Millennium simulation though.

Our primary constraints include galaxy mass functions for each of stars (already presented in Fig.~\ref{fig:smf}), H\,\textsc{i}, and H$_2$ at $z=0$, in addition to the H\,\textsc{i}--stellar mass scaling relation.  The mass function, $\Phi$, describes the number density of galaxies with a particular mass (in a particular baryonic species) per unit volume.  Of these, only the stellar mass function was originally used to constrain \textsc{sage}.  This was primarily because that model did not break cold gas into H\,\textsc{i} and H$_2$ \citep[for a version of \textsc{sage} that did do this, see][]{wolz16}.  

\begin{figure}
	\centering
	\includegraphics[width=0.95\textwidth]{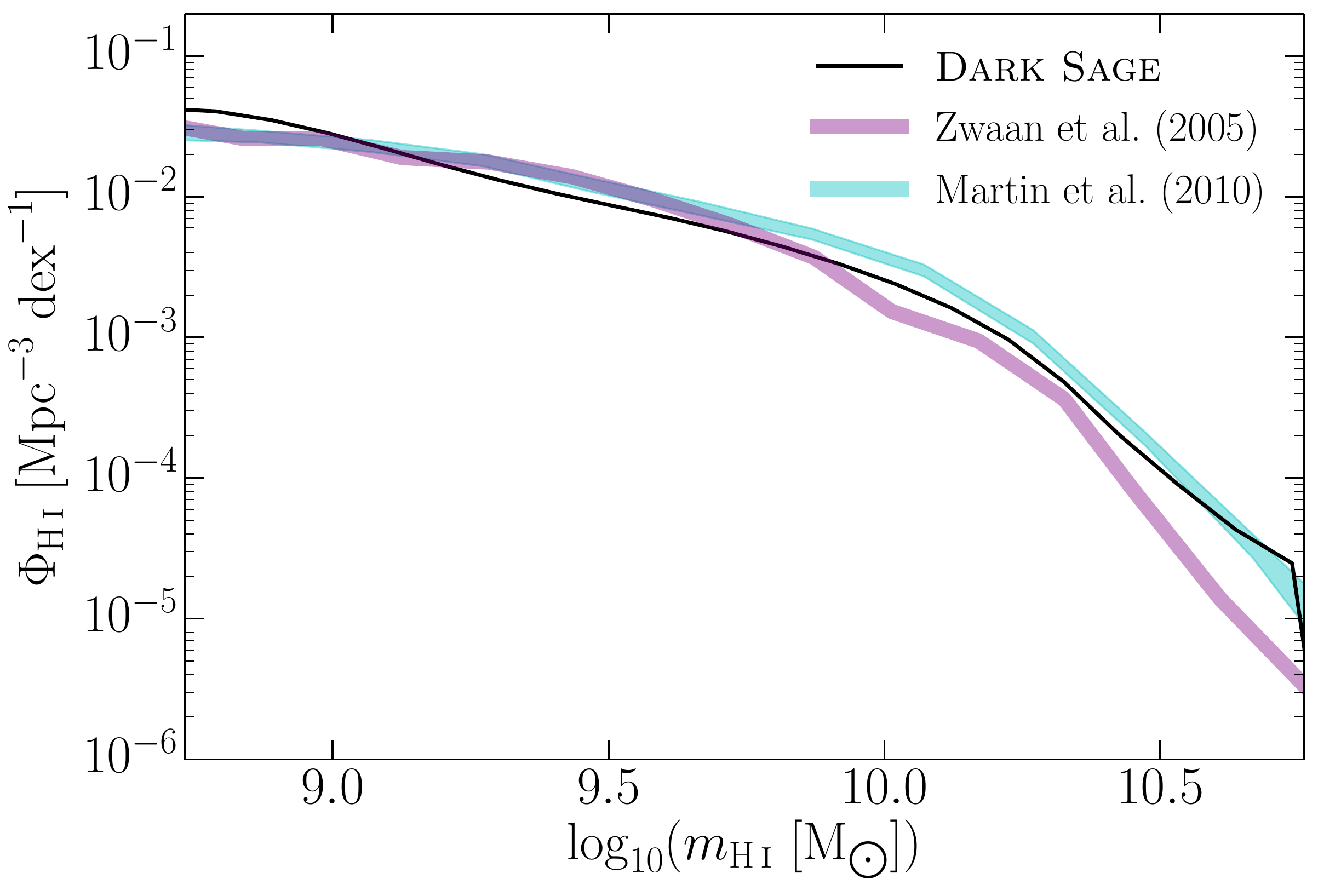}
	\caption{H\,\textsc{i} mass function of \textsc{Dark Sage} galaxies at $z=0$ compared to observations \citep{zwaan05,martin10}.}
	\label{fig:himf}
\end{figure}

We constrain the H\,\textsc{i} mass function using observational data from \citet{zwaan05}, who built their mass function from the H\,\textsc{i} Parkes All Sky Survey.  This is shown in Fig.~\ref{fig:himf}.  Each mass bin had a sufficient number of galaxies such that their published Poisson errors are smaller than the thickness of the line over the range plotted here.  For completeness, we also compare the mass function measured equivalently from the Arecibo Legacy Fast ALFA survey \citep{martin10}, covering the published uncertainty range.

\begin{figure}
	\centering
	\includegraphics[width=0.95\textwidth]{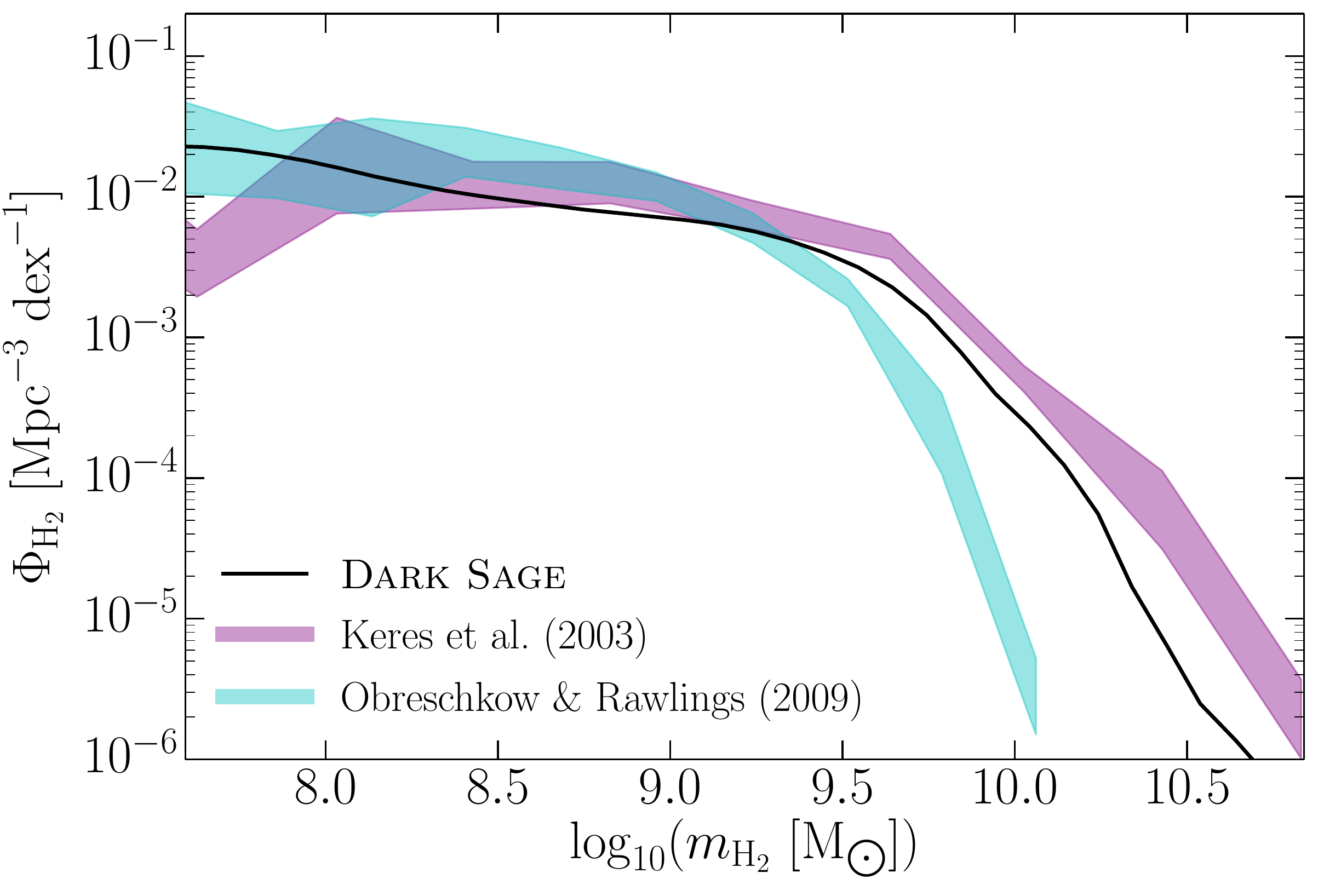}
	\caption{H$_2$ mass function of \textsc{Dark Sage} galaxies at $z=0$ compared to observations \citep{keres03,obreschkow09}.}
	\label{fig:h2mf}
\end{figure}

To constrain the H$_2$ mass function, we initially used data from \citet{keres03}, as shown in Fig.~\ref{fig:h2mf}.  Those authors measured the CO luminosity function from observations of local galaxies, where the derived H$_2$ mass function assumed a constant conversion factor between the CO and H$_2$ content.  In the later stages of calibration, we found the model that best fit the other constraints fell in between the H$_2$ mass functions of \citet{keres03} and \citet{obreschkow09}.  The latter authors used the same data but applied a variable conversion factor from CO to H$_2$. 

\begin{figure}
	\centering
	\includegraphics[width=0.95\textwidth]{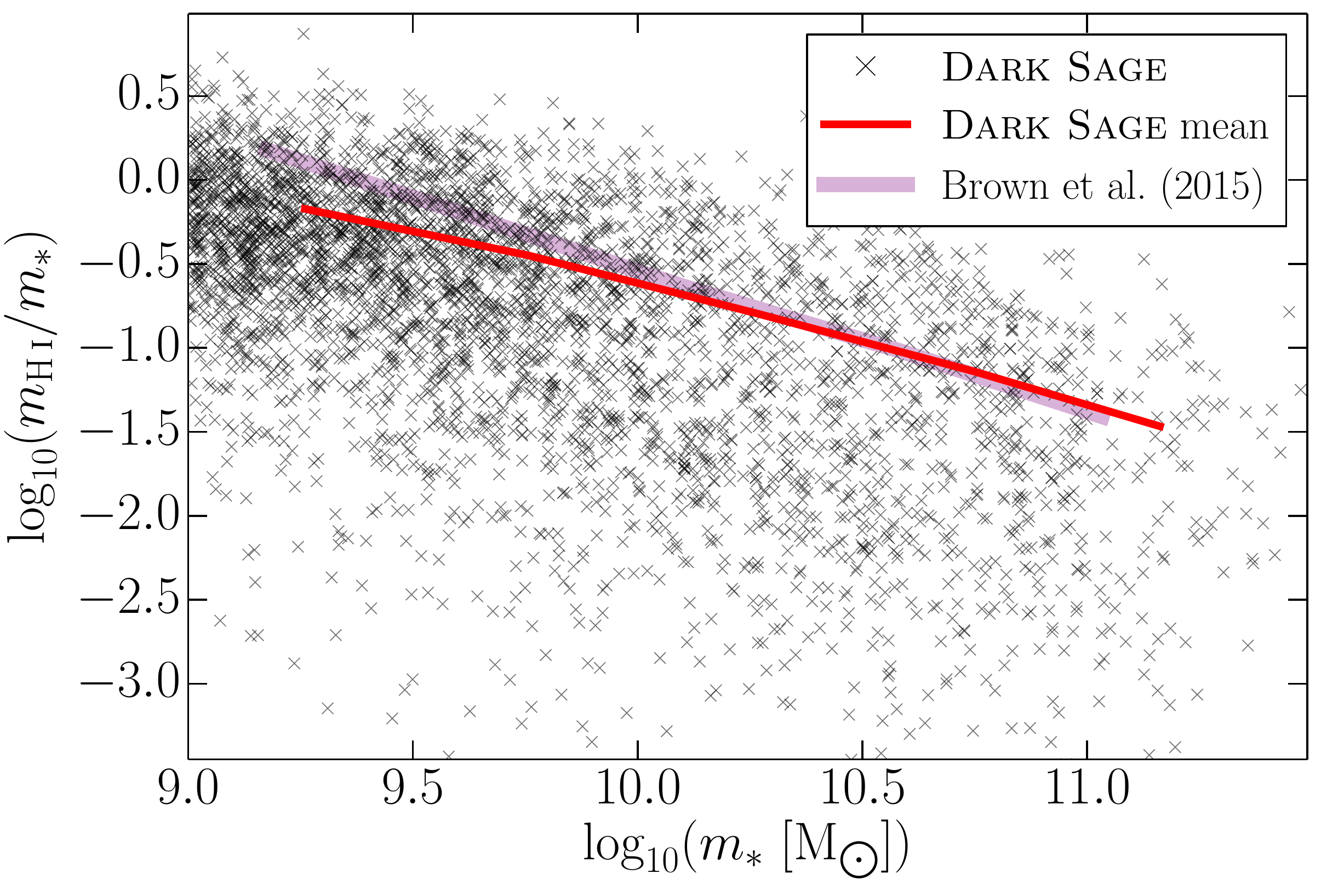}
	\caption{H\,\textsc{i} mass fraction as a function of stellar mass at $z=0$.  We show 4000 representative galaxies from \textsc{Dark Sage} as well as the mean trend to be compared against the observed relation of \citet{brown15}.}
	\label{fig:hifrac}
\end{figure}

Through stacking H\,\textsc{i} observations of galaxies which would otherwise be non-detections, \citet{brown15} found an average scaling relation between the H\,\textsc{i} mass fraction of galaxies and their stellar mass.  We use this to constrain the H\,\textsc{i} mass fractions of \textsc{Dark Sage} galaxies.  We show these in Fig.~\ref{fig:hifrac} and include the mean trend, binned similarly to the observational data.  Observational errors from jackknifing are smaller than the width of the line.  We note that the lowest mass bin, $10^9 \lesssim m_* / \mathrm{M}_{\odot} \lesssim 10^{9.5}$, always showed a deficit in the H\,\textsc{i} fraction throughout our calibration process.

\begin{figure}
	\centering
	\includegraphics[width=0.95\textwidth]{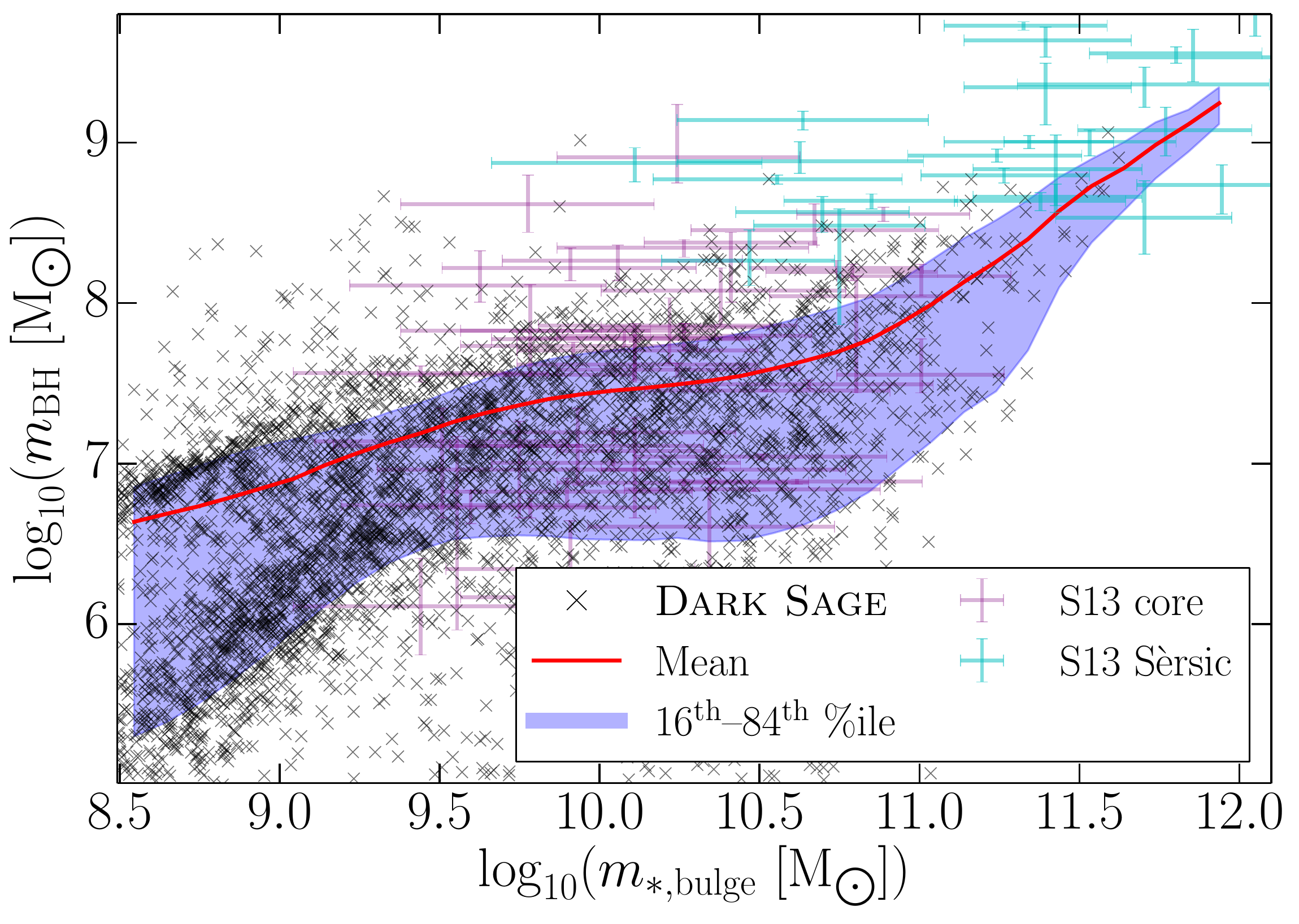}
	\caption{Black hole--bulge mass relation at $z=0$.  We show 4000 representative galaxies from \textsc{Dark Sage}, and include the mean and 16$^{\rm th}$--84$^{\rm th}$ percentile range of the relation.  Observational data from \citet{scott13} are compared.}
	\label{fig:bhbulge}
\end{figure}

\begin{figure}
	\centering
	\includegraphics[width=0.95\textwidth]{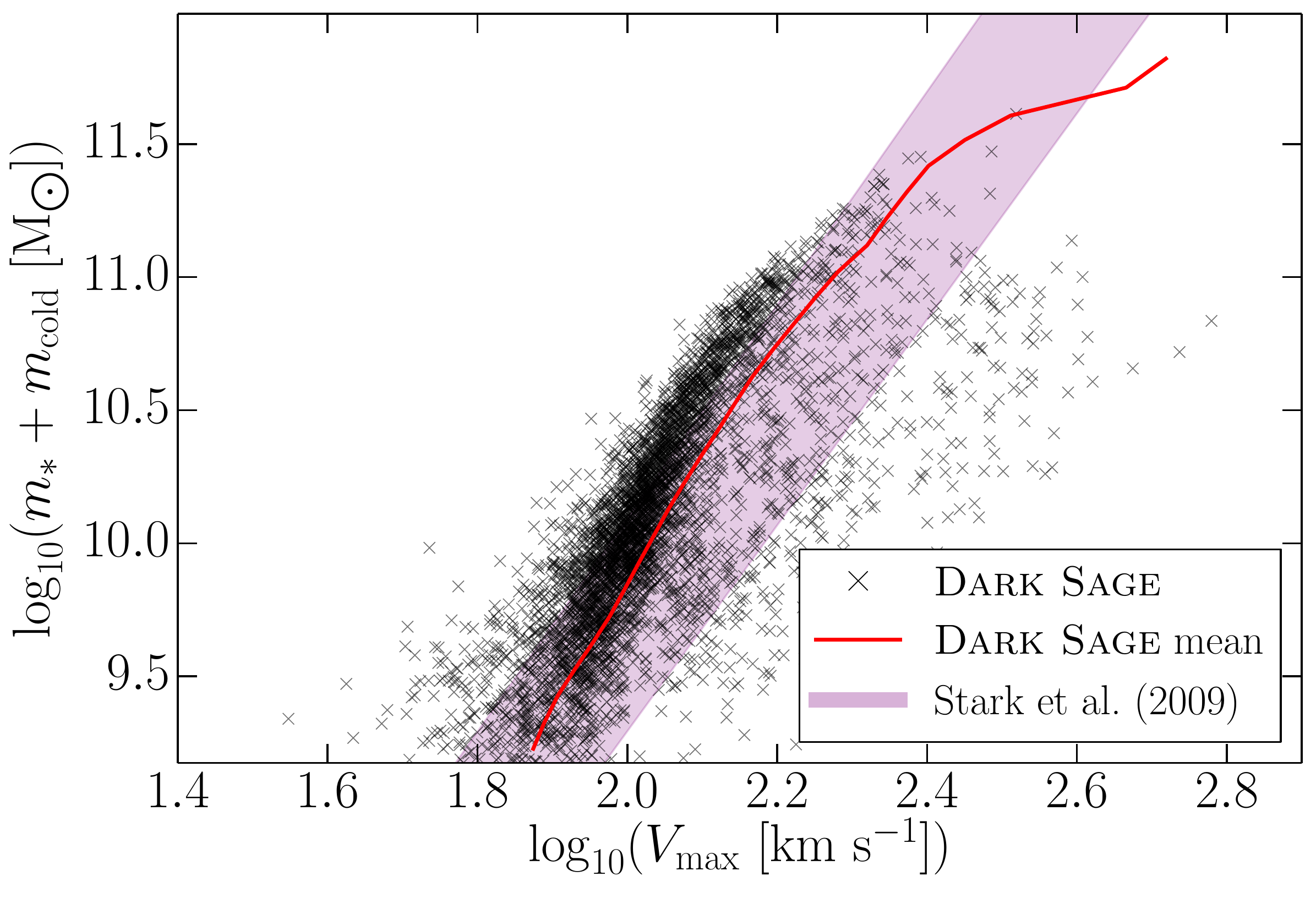}
	\caption{Baryonic Tully--Fisher relation for \textsc{Dark Sage} galaxies at $z=0$ with bulge-to-total ratios between 0.1 and 0.5.  We show 4000 example galaxies in the axes and the mean rotation velocity for mass bins of width 0.1 dex.  Compared is the observational trend published by \citet{stark09}.}
	\label{fig:btf}
\end{figure}

\begin{figure}
	\centering
	\includegraphics[width=0.95\textwidth]{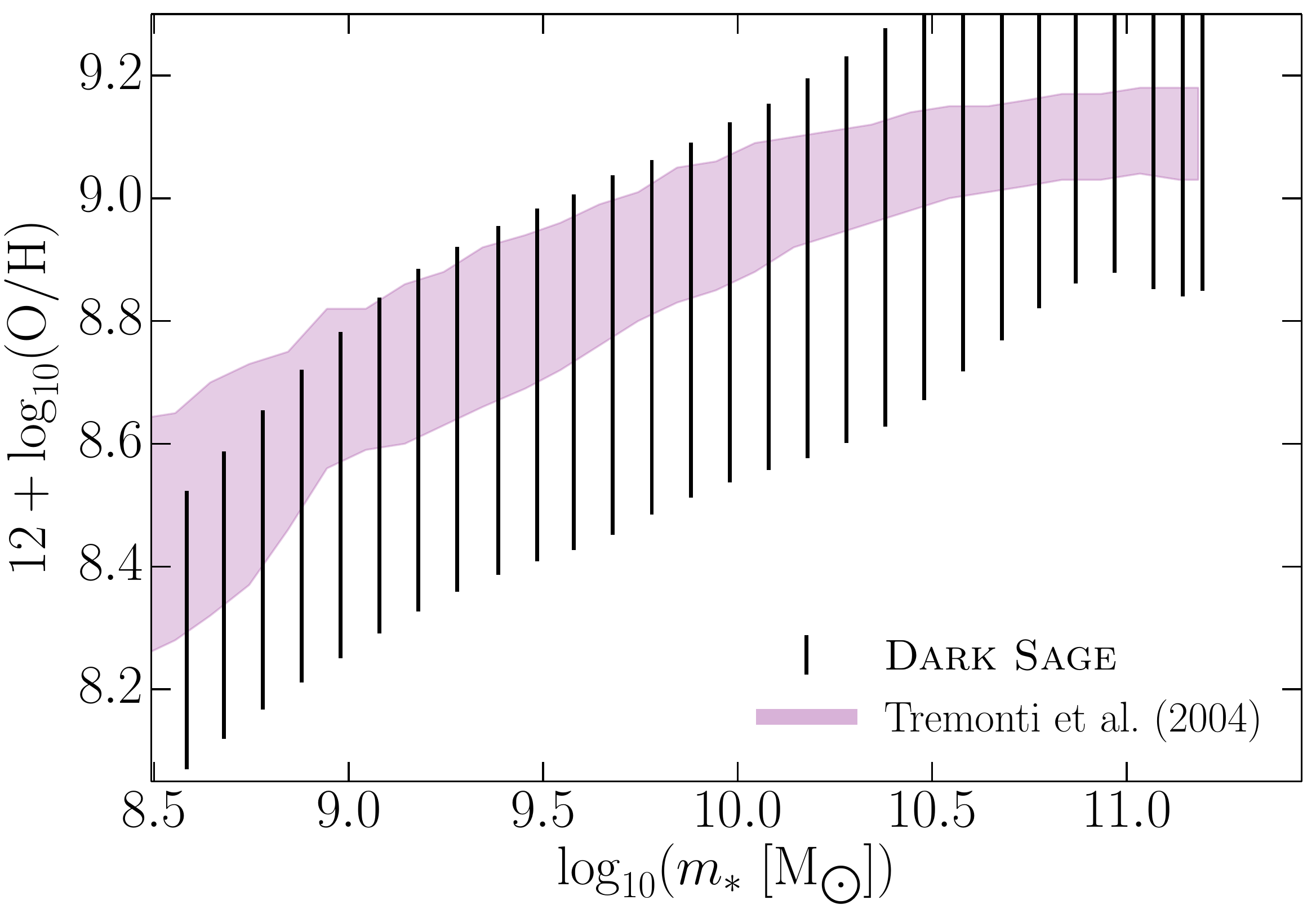}
	\caption{Stellar mass--gas metallicity relation at $z=0$.  Bars cover the 16$^{\rm th}$--84$^{\rm th}$ percentile range of range for \textsc{Dark Sage} galaxies in each bin, matched closely to the bins used in the observed data of \citet{tremonti04}.}
	\label{fig:massmet}
\end{figure}

The remaining constraints were identical to what was done for \textsc{sage}.  Fig.~\ref{fig:bhbulge} provides points and the sequence for the black hole--bulge mass relation of \textsc{Dark Sage} galaxies.  We compare against observational data with uncertainties from \citet{scott13}, who split galaxies into S\`{e}rsic and core-S\`{e}rsic types.
Fig.~\ref{fig:btf} shows the Baryonic Tully--Fisher relation (maximum rotational velocity against the sum of stellar and cold gas mass) for relevant \textsc{Dark Sage} galaxies, with the fitted relation to observational data from \citet{stark09} including their random uncertainties.
Fig.~\ref{fig:massmet} displays the stellar mass--gas metallicity relation, by calculating $12 + \log_{10}({\rm O/H}) = 9 + \log_{10}(m_{Z,\mathrm{cold}} / 0.02\, m_{\rm cold})$.  Both \textsc{Dark Sage} and the observational data \citep{tremonti04} display the 16$^{\rm th}$--84$^{\rm th}$ percentile range of metallicity from bins of approximately 0.1-dex width in stellar mass.

\begin{figure}
	\centering
	\includegraphics[width=0.95\textwidth]{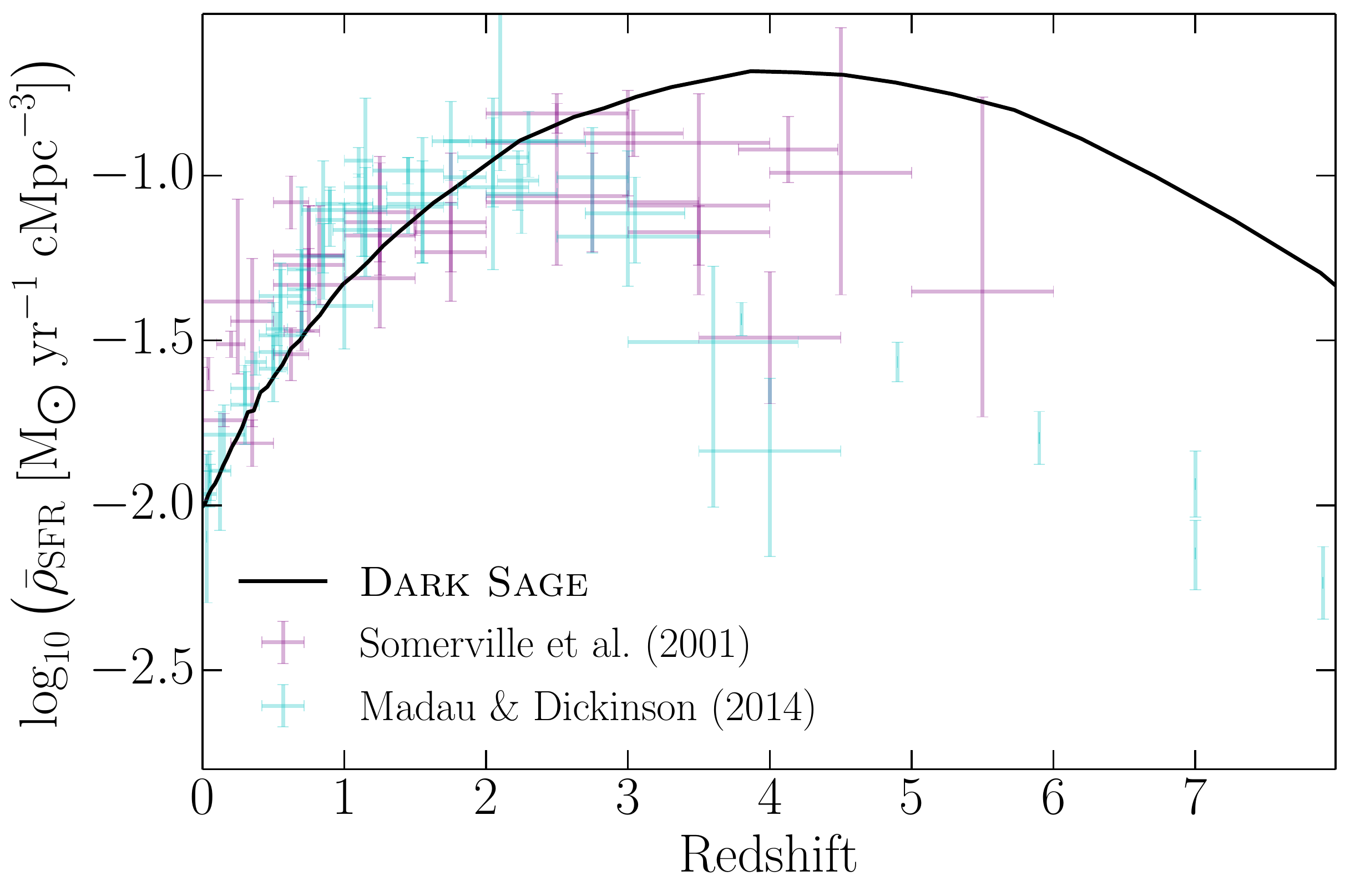}
	\caption{Madau--Lilly diagram.  Observational data originate from a variety of sources; we constrained \textsc{Dark Sage} against those compiled by \citet{somerville01}.  Compared are data compiled by \citet{madau14}.}
	\label{fig:sfrd}
\end{figure}

Finally, we present the comoving star formation rate density history, also known as the Madau--Lilly diagram \citep[\`{a} la][]{lilly96,madau96}, in Fig.~\ref{fig:sfrd}.  Observational data used for this constraint were compiled by \citet{somerville01}, with a complete set of references in their table A2.  We also compare the more recent compilation presented by \citet[][see their table 1]{madau14}, modified for a \citet{chabrier03} initial mass function.  These newer data suggest we are overproducing stars at $z \gtrsim 3$.
	

%% file: AppA.tex
\section{Rotation curves}
\label{app:rotcurves}

It is important for many parts of the model to convert between $j$ and $r$ for the disc annuli.  From equation (\ref{eq:j}), this means $v_{\rm circ}(r)$ must be known.  At each time-step, we construct a rotation curve for each disc, where
\begin{equation}
v_{\rm circ}(r) = \sqrt{\frac{G\, M(<r)}{r}}~.
\end{equation}
We then iteratively solve the equation
\begin{equation}
j = r\,v_{\rm circ}(r) = \sqrt{G\, M(<r)\, r}
\label{eq:j2r}
\end{equation}
to calculate a radius for each annulus edge.  The mass internal to radius $r$, $M(<r)$, is determined by summing the mass profiles of each of the components in the (sub)halo,
\begin{multline}
M(<r) = m_{\rm cold}(<j) + m_{*,\mathrm{disc}}(<j) + m_{\rm DM}(<r) + m_{\rm hot}(<r)\\ + m_{\rm m-bulge}(<r) + m_{\rm i-bulge}(<r) +  m_{\rm ICS}(<r) + m_{\rm BH}~.
\end{multline}
The mass internal to $j$ in the disc and the black hole mass, $m_{\rm BH}$, is directly known, while we describe how each of the other components is modelled below.  

The value of $v_{\rm circ}(r)$ is not allowed to exceed the $V_{\rm max}$ of the halo as recorded by the halo finder.  This ensures rotation curves do not spike to excessively large values at low radius, which happens for disc-dominated systems without the restriction.  This problem is associated with the excessive pooling of baryonic mass towards the centres of discs, shared by many models of disc evolution \citep[see][and references therein]{stringer07}.

\subsection{Dark matter}
We use the NFW profile \citep{nfw96,nfw97} for dark-matter haloes, which gives a cumulative mass of
\begin{multline}
m_{\rm DM}(<r) = m_{\rm DM} \left[ \ln\left(\frac{r+r_h}{r_h}\right) - \frac{r}{r+r_h} \right] \\ \left[ \ln\left(\frac{R_{\rm vir}+r_h}{r_h}\right) - \frac{R_{\rm vir}}{R_{\rm vir}+r_h} \right]^{-1}~,
\end{multline}
where $r_h$ describes a scale radius, related to the concentration of the halo,
\begin{equation}
c \equiv R_{\rm vir} / r_h~.
\end{equation}

Over cosmic time, as haloes accumulate mass and condense, their concentrations increase.  Larger haloes also achieve higher concentrations in shorter time-scales.  Using a series of $N$-body simulations which covered five orders of magnitude of halo mass, \citet{dutton14} showed there is a generic dependence of the concentration of dark-matter-only haloes on their mass and redshift:
\begin{subequations}
\label{eq:c_DM}
\begin{equation}
\log_{10} (c_{\rm DM}) = a + b \log_{10}(M_{\rm vir} h / 10^{12} \mathrm{M}_{\odot})~,
\end{equation}
\begin{equation}
a = 0.520 + 0.385 e^{-0.617 z^{1.21}}~,
\end{equation}
\begin{equation}
b = -1.01 + 0.026 z~.
\end{equation}
\end{subequations}
This expression is valid for $z<5$.  For higher redshifts, we approximate the concentration of haloes to be as if it were redshift 5.  This has a negligible effect on the evolution of galaxies in \textsc{Dark Sage}, given the relatively few snapshots considered at $z>5$ compared to those at lower redshift.

It is now widely accepted that baryonic physics plays an important role in the distribution of dark matter in haloes, especially in their centres \citep[see, e.g.,][]{duffy10,dicintio14,brook15,schaller15}.  With hydrodynamic simulations, \citet{dicintio14} showed that the effect of baryons on the concentration of dark-matter haloes can be captured by the proportion of the virial mass in the form of stars in the galaxy:
\begin{equation}
c = \left\{1 + 3 \times 10^{-5}\, e^{3.4[\log_{10}(m_* / M_{\rm vir}) + 4.5]} \right\} c_{\rm DM}~.
\label{eq:c_SPH}
\end{equation}
We substitute equation (\ref{eq:c_DM}) into (\ref{eq:c_SPH}) to solve for the dark-matter halo concentration in \textsc{Dark Sage}.

In truth, an NFW profile is insufficient to account for all the baryonic effects on the dark-matter profile.  A more generic profile for the family of \citet{jaffe83}, \citet{hernquist90}, and NFW profiles \citep[\`{a} la][]{merritt06} would be more accurate, where \citet{dicintio14} provide scalings for the parameters of these profiles. Solving equation (\ref{eq:j2r}) becomes immensely more computationally expensive, however, and this level of detail is beyond that necessary for \textsc{Dark Sage}.

\subsection{Hot gas}
The cumulative mass profile of hot gas in \textsc{Dark Sage} comes directly from the assumption that the gas is a singular isothermal sphere:
\begin{equation}
m_{\rm hot}(<r) = m_{\rm hot} \frac{r}{R_{\rm vir}}~.
\end{equation}

\subsection{Bulges and spheroids}
We consider merger-driven and instability-driven bulges each as independent components of a galaxy, but approximate both as having \citet{hernquist90} profiles truncated at the virial radius,
\begin{equation}
m_{\rm bulge}(<r) = m_{\rm bulge} \left[ \frac{r(R_{\rm vir}+a_{\rm bulge})}{R_{\rm vir}(r+a_{\rm bulge})} \right]^2~,
\label{eq:hernquist}
\end{equation}
where the half-mass radius of the bulge component is $(1+\sqrt{2})a_{\rm bulge}$, assuming $a_{\rm bulge} \ll R_{\rm vir}$.  For the merger-driven bulge, we use the empirical average size--mass scaling from \citet{sofue16},
\begin{equation}
\log_{10}\left(\frac{a_{\rm m-bulge}}{\mathrm{kpc}}\right) = \frac{1}{1.13} \left[\log_{10}\left(\frac{m_{\rm m-bulge}}{\mathrm{M}_{\odot}}\right) - 10.21 \right]~.
\end{equation}
For the instability-driven bulge, we use the observed average size--mass scaling of pseudobulges from \citet{fisher08},
\begin{equation}
a_{\rm i-bulge} = 0.2\, r_d\, (1 + \sqrt{2})^{-1}~,
\end{equation}
where $r_d$ is taken from equation (\ref{eq:r_d}).  

This model for the bulge size is incredibly crude, but its precision is not an important factor for the goals of the model in studying discs.  For a modified version of \textsc{sage} with a thorough evolution of bulge sizes, see \citet{tonini16}.

\subsection{Intracluster stars}
Modelling the mass distribution of intracluster stars is only of moderate significance for brightest cluster galaxies.  \citet*{gonzalez05} found the scale radius for intracluster light correlates with that of the elliptical brightest cluster galaxy.  We follow their simple scaling relation, assuming intracluster stars are distributed in a \citet{hernquist90} sphere (equation \ref{eq:hernquist}) with
\begin{equation}
a_{\rm ICS} = 13\, a_{\rm m-bulge}~.
\end{equation}